# Cooperative Relay Broadcast Channels [1] [2]

Yingbin Liang and Venugopal V. Veeravalli [3]

**Abstract**

The capacity regions are investigated for two relay broadcast channels (RBCs), where relay links are incorporated into standard two-user broadcast channels to support user cooperation. In the first channel, the *Partially Cooperative Relay Broadcast Channel*, only one user in the system can act as a relay and transmit to the other user through a relay link. An achievable rate region is derived based on the relay using the decode-and-forward scheme. An outer bound on the capacity region is derived and is shown to be tighter than the cut-set bound. For the special case where the Partially Cooperative RBC is degraded, the achievable rate region is shown to be tight and provides the capacity region. Two Gaussian cases of the Partially Cooperative RBC are studied. For the system where the additive Gaussian noise term at one receiver is a degraded version of the other, which we refer to as the D-AWGN Partially Cooperative RBC, the capacity region is established. For the system where the additive Gaussian noise term at one receiver is independent of the other, which we refer to as the AWGN Partially Cooperative RBC, inner and outer bounds on the capacity region are derived and are shown to be close. Furthermore, it is shown that feedback does not increase the capacity region for the degraded Partially Cooperative RBC, but that it may improve the capacity region for the non-degraded Partially Cooperative RBC. In particular, feedback improves the capacity region for the AWGN Partially Cooperative RBC. In the second channel model being studied in the paper, the *Fully Cooperative Relay Broadcast Channel*, both users can act as relay nodes and transmit to each other through relay links. This is a more general model than the Partially Cooperative RBC. All the results for Partially Cooperative RBCs are correspondingly generalized to the Fully Cooperative RBCs. In particular, the capacity regions are established for the degraded Fully Cooperative RBC and its Gaussian example of the D-AWGN Fully Cooperative RBC. The capacity region is also established for the Fully Cooperative RBC with feedback. It is further shown that the AWGN Fully Cooperative RBC has a larger achievable rate region than the AWGN Partially Cooperative RBC. The results illustrate that relaying and user cooperation are powerful techniques in improving the capacity of broadcast channels.

## 1 Introduction

Cooperative relaying of information between users is emerging as a powerful technique for improving the reliability and throughput of wireless networks. The building block of such relay networks, the three-terminal relay channel, was first introduced by van der Meulen [46], and was comprehensively studied by Cover and El Gamal [5]. Recently, this channel has been further studied in a variety of contexts including Gaussian relay channels (e.g., [13, 22]), fading relay channels (e.g., [43, 44, 24, 47, 18, 15, 50, 30]), relay channels with complexity constraints [34], relay channels with multiple antennas (e.g.,[24, 47]), and relay channels with orthogonal components (e.g., [26, 18, 12, 21, 35, 23, 28]). More complicated relay networks have also been studied including relay networks with multiple relay nodes simultaneously relaying information to the destination (e.g.,[24, 14, 42, 32, 33]), relay networks with multiple levels of relay nodes forwarding information from one level to next (e.g.,[16, 49, 48, 24, 39]), and relay networks with multiple cooperative sources or destinations (e.g., [16, 17, 19, 20]). Furthermore, these information-theoretic studies of relay networks have motivated practical relaying protocol and coding design to achieve user-cooperative diversity (e.g.,[27, 35, 45, 38, 1]).

---

[1]The material in this paper was presented in part in the IEEE International Symposium on Information Theory, Chicago, Illinois in 2004, and in part in the Wirelesscom, Symposium on Information Theory, Maui, Hawaii in 2005.

[2]This research was supported by the NSF CAREER/PECASE award CCF 0049089 and by a Vodafone Foundation Graduate Fellowship.

[3]The authors are with the Department of Electrical and Computer Engineering and the Coordinated Science Laboratory, University of Illinois at Urbana-Champaign, Urbana IL 61801; e-mail: {yliang1,vvv}@uiuc.edu

For centralized networks, to date much of work on this topic has focused on the uplink (from the users to a base station or access point). Cooperative diversity schemes, where one user may share another user's resources to improve its transmission rate, have been explored in a number of recent works (see, e.g. [43, 44, 26]). The use of a relay node to assist all the users in a multi-access channel has been studied in [25, 24, 40, 41], and bounds on the corresponding capacity region have been derived.

In this paper, we study the impact of relaying and user cooperation on downlink systems. Since in centralized wireless networks such as cellular and WiFi data networks, mobile users have been demanding increasingly higher data rates on the downlink, improving the throughput on the downlink has become an extremely important issue. This application motivates us to study the downlink or broadcast channel that exploits the technique of relaying and user cooperation to achieve higher throughput. We introduce and study two such systems of relay broadcast channels (RBCs), where relay links are incorporated into the standard broadcast channel [3, 4] to assist broadcast transmission. These RBC models represent the most fundamental user-cooperative downlink systems and capture the essential roles of user cooperation in downlink communications. We focus on the two-user case of these RBCs, and study the gains in capacity region offered by relaying and user cooperation.

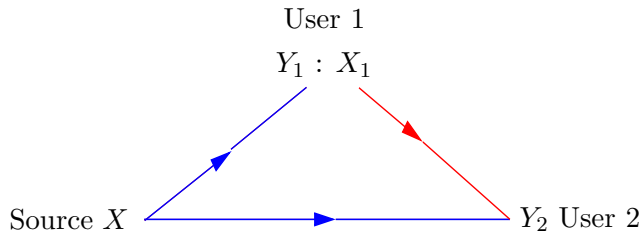

Figure 1: Partially Cooperative Relay Broadcast Channel

We first study the *Partially Cooperative Relay Broadcast Channel*, which is based on the standard two-user broadcast channel with one source attempting to transmit both common information and private information sets to two users. Moreover, user 1 acts as a standard relay node [46, 5] and transmits cooperative information to user 2 through a relay link (see Figure 1). A possible motivation for studying this channel is that in a two-user broadcast system usually one user (denoted by user 1) has a "better" channel from the source than the other user (denoted by user 2), and hence user 1 may decode the information intended for user 2 in addition to its own information. Then user 2 should benefit from having a relay link from user 1. Such user cooperation is particularly useful when a mobile user experiences a deep fading state and can still maintain a reliable communication with the help of another user (the relay node).

We further study the Partially Cooperative RBC with feedback, where, similarly as the relay channel with feedback [5, Section V], the outputs at user 2 are provided to user 1 and the outputs at both users 1 and 2 are provided to the source all through perfect feedback links. Our motivation to study this feedback channel is not only because feedback is a natural topic in the study of broadcast channels [10, 11, 36] and relay channels [5], but also because study of feedback channels provides the insight on what information is useful for user cooperation. Furthermore, the proof of the converse for the capacity region of a feedback channel suggests a way to obtain a tighter outer bound than the cut-set bound on the capacity region for the corresponding relay broadcast channel without feedback.

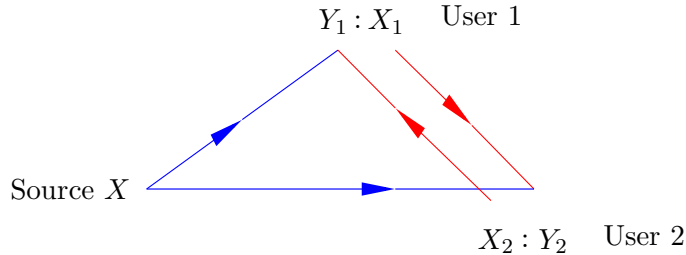

Figure 2: Fully Cooperative Relay Broadcast Channel

We then move on to study a more general model, the *Fully Cooperative Relay Broadcast Channel*, where both users can transmit cooperative information to each other through relay links (see Figure 2). In this channel, both users can potentially gain in capacity due to this cooperative relaying, which will be demonstrated in the paper by a Gaussian example. We also study the Fully Cooperative RBC with feedback and derive results that are parallel to those for the Partially Cooperative RBC.

For both the Partially Cooperative RBC and the Fully Cooperative RBC, we illustrate our results via two Gaussian cases. The first case is the D-AWGN Partially/Fully Cooperative RBC, where the outputs at the two receivers are corrupted by (physically) degraded Gaussian noise terms, i.e., if the random noise terms are denoted by $Z_1$ and $Z_2$, then $Z_2 = Z_1 + Z'$ and $Z_1$ and $Z'$ are independent Gaussian random variables. Such degraded channel models are of information-theoretic interest, because we can usually establish the capacity regions for these models. Furthermore, the proofs of the converse for the capacity regions of these degraded models suggest techniques for obtaining outer bounds on the capacity regions of non-degraded Gaussian channels. The second Gaussian case being studied in this paper is the AWGN Partially/Fully Cooperative RBC, where the outputs at the two receivers are corrupted by independent Gaussian noise terms. This Gaussian case represents a natural channel encountered in practice.

We note that a model of the *broadcast channel with cooperating receivers* has been introduced and studied in [9, 8]. This model differs from the Partially/Fully Cooperative RBC being studied in this paper in that the relay links for user cooperation are orthogonal to the original broadcast channel. We also note that a related relay broadcast channel model, where an additional relay node is introduced to broadcast systems to assist all users, has been introduced and studied in [24, 29].

In the following, we summarize the main results of this paper.

For the discrete memoryless Partially and Fully Cooperative RBCs:

- We derive achievable rate regions, i.e., inner bounds on the capacity regions;
- We provide outer bounds on the capacity regions, and show that these outer bounds are tighter than the cut-set bounds;

- We establish the capacity regions for the degraded Partially and Fully Cooperative RBCs, where the previous inner and outer bounds match;

- We establish the capacity regions for the Partially and Fully Cooperative RBCs with feedback. We show that feedback does not increase the capacity regions for the degraded Partially and Fully Cooperative RBCs, but that it may improve the capacity regions for the non-degraded channels. In particular, feedback improves the capacity regions for the AWGN Partially and Fully Cooperative RBCs.

For the Gaussian Partially and Fully Cooperative RBCs, we summarize our results in the following diagram, where $\mathcal{C}(\cdot)$ denotes the capacity region and $\mathcal{R}(\cdot)$ denotes the achievable rate region. We assume the Gaussian noise variance at user 1 is less than the Gaussian noise variance at user 2.

$$\boxed{\begin{array}{l} \mathcal{C}(\text{D-AWGN Partially Cooperative RBC}) \\ = \mathcal{C}(\text{D-AWGN Partially Cooperative RBC with Feedback}) \\ = \mathcal{C}(\text{D-AWGN Fully Cooperative RBC}) \\ = \mathcal{C}(\text{D-AWGN Fully Cooperative RBC with Feedback}) \end{array}}$$

$$= \quad \mathcal{R}(\text{AWGN Partially Cooperative RBC})$$

$$\subset \quad \mathcal{R}(\text{AWGN Fully Cooperative RBC})$$

$$\subset \quad \boxed{\begin{array}{l} \mathcal{C}(\text{AWGN Partially Cooperative RBC with Feedback}) \\ = \mathcal{C}(\text{AWGN Fully Cooperative RBC with Feedback}) \end{array}}$$

The notation in this paper mainly follows the following rules. Upper case letters indicate random variables, and lower case letters indicate deterministic variables or realizations of the corresponding random variables. There are some exceptions, but these will be clarified where they appear in the paper. We use $\underline{x}$ or $x^n$ to indicate the vector $(x_1, \ldots, x_n)$, and use $x_i^n$ to indicate the vector $(x_i, \ldots, x_n)$. We define two functions: $\bar{x} := 1 - x$ and $\mathcal{C}(x) := \frac{1}{2}\log(1+x)$. Throughout the paper, the logarithmic function is to the base 2.

In the following sections, we first present the results for the Partially Cooperative RBC and the results for this channel with feedback. We then present the results for the Fully Cooperative RBC and the results for this channel with feedback. We finally discuss and compare the achievable regions for the AWGN case of these RBCs with all nodes being subject to a sum power constraint. We end the paper with some concluding remarks.

## 2 Partially Cooperative Relay Broadcast Channels

In this section, we first introduce the channel model for the Partially Cooperative RBC, and then present the main results. We further illustrate the results via two Gaussian channel examples.

### 2.1 System Model

To define the Partially Cooperative RBC, we use $x$ to denote the source input, $x_1$ to denote the relay input of user 1, and $y_1$ and $y_2$ to denote the outputs at user 1 and user 2, respectively.

**Definition 1.** A Partially Cooperative RBC consists of a channel input alphabet $\mathcal{X}$, a relay input alphabet $\mathcal{X}_1$, two channel output alphabets $\mathcal{Y}_1$ and $\mathcal{Y}_2$, and a probability transition function $p(y_1, y_2 | x, x_1)$ (see Figure 3).

Note that the channel input-output relationship is similar to that of the standard relay channel [5], but now the relay node (user 1) also has its own message to decode. We assume throughout the paper that the channel is memoryless.

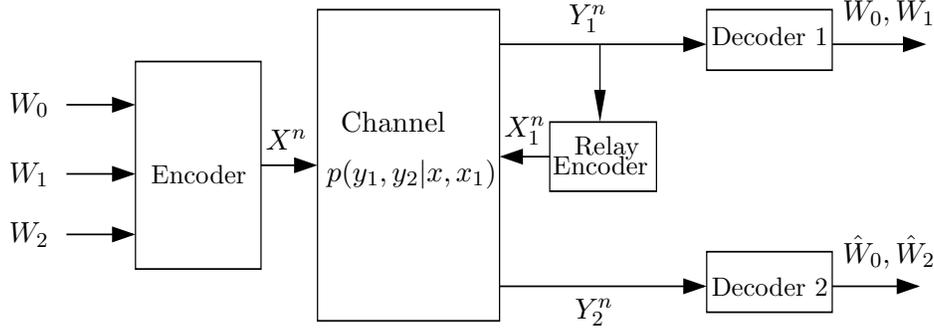

Figure 3: Partially Cooperative Relay Broadcast Channel

**Definition 2.** A $\left(2^{nR_0}, 2^{nR_1}, 2^{nR_2}, n\right)$ code for a Partially Cooperative RBC consists of:

- Three message sets: $\mathcal{W}_0 = \{1, 2, \ldots, 2^{nR_0}\}$, $\mathcal{W}_1 = \{1, 2, \ldots, 2^{nR_1}\}$ and $\mathcal{W}_2 = \{1, 2, \ldots, 2^{nR_2}\}$;

- An encoder: $\mathcal{W}_0 \times \mathcal{W}_1 \times \mathcal{W}_2 \to \mathcal{X}^n$, which maps each message tuple $(W_0, W_1, W_2)$ to a codeword $x^n \in \mathcal{X}^n$;

- A set of relay functions $\{f_i\}_{i=1}^n$ such that
$$x_{1,i} = f_i(y_{1,1}, \ldots, y_{1,i-1}), \qquad 1 \leq i \leq n;$$

- Two decoders: one at user 1, $\mathcal{Y}_1^n \to \mathcal{W}_0 \times \mathcal{W}_1$, which maps a received sequence $y_1^n$ to a message pair $(W_0, W_1) \in \mathcal{W}_0 \times \mathcal{W}_1$; and the other at user 2, $\mathcal{Y}_2^n \to \mathcal{W}_0 \times \mathcal{W}_2$, which maps $y_2^n$ to a message pair $(W_0, W_2) \in \mathcal{W}_0 \times \mathcal{W}_2$.

We note that in the above definition, $W_0$ indicates the common message that needs to be decoded at both users, and $W_1$ and $W_2$ are private messages that need to be decoded at users 1 and 2, respectively.

The probability of error when the message tuple $(W_0, W_1, W_2)$ is sent is defined as
$$P_e^{(n)}(W_0, W_1, W_2) = P\left((\hat{W}_0, \hat{W}_1) \neq (W_0, W_1) \text{ or } (\hat{W}_0, \hat{W}_2) \neq (W_0, W_2)\right), \qquad (1)$$

and the average probability of error is defined by assuming that the message $(W_0, W_1, W_2)$ is uniformly distributed over $\mathcal{W}_0 \times \mathcal{W}_1 \times \mathcal{W}_2$ and is given by
$$P_e^{(n)} = \frac{1}{2^{nR_0} 2^{nR_1} 2^{nR_2}} \sum_{W_0=1}^{2^{nR_0}} \sum_{W_1=1}^{2^{nR_1}} \sum_{W_2=1}^{2^{nR_2}} P_e^{(n)}(W_0, W_1, W_2). \qquad (2)$$

The rate tuple $(R_0, R_1, R_2)$ is said to be achievable for the Partially Cooperative RBC if there exists a sequence of $\left(2^{nR_0}, 2^{nR_1}, 2^{nR_2}, n\right)$ codes with average error probability $P_e^{(n)} \to 0$ as $n$ goes to infinity.

**Definition 3.** *A Partially Cooperative RBC is* degraded *if the transition probability satisfies*

$$p(y_1, y_2|x, x_1) = p(y_1|x, x_1)p(y_2|y_1, x_1). \tag{3}$$

*i.e., $y_2$ is independent of $x$, conditioned on $y_1$ and $x_1$.*

## 2.2 Discrete Memoryless Partially Cooperative RBCs

A motivation to study the Partially Cooperative RBC is that in many cases one user in the broadcast channel has the capability to decode at higher rate than the other user. Hence this user may also decode the message for the other user in addition to the message for itself, and then forward this information for the other user. The following achievable rate region for the Partially Cooperative RBC is based on this idea where the relay (user 1) employs the decode-and-forward relaying scheme to help user 2.

**Theorem 1.** *(Inner bound on Capacity Region for Partially Cooperative RBC) A rate tuple $(R_0, R_1, R_2)$ is achievable for the discrete memoryless Partially Cooperative RBC if*

$$\begin{aligned} R_0 + R_2 &< \min\left\{I(U, X_1; Y_2), I(U; Y_1 \mid X_1)\right\}, \\ R_1 &< I(X; Y_1 \mid U, X_1) \end{aligned} \tag{4}$$

*for some joint distribution $p(u, x_1, x)p(y_1, y_2|x, x_1)$, where $U$ is bounded in cardinality by $|\mathcal{U}| \leq |\mathcal{X}| \cdot |\mathcal{X}_1| + 2$.*

*Proof.* See Appendix A for an outline of the proof. □

The achievable region based on the decode-and-forward scheme in Theorem 1 serves as an example to show that relaying from user 1 to user 2 indeed helps to enlarge the capacity region of the original broadcast channel. This will be further demonstrated by Gaussian channels later. More importantly, we will show that this achievable region is tight for the special case of degraded RBCs, and we hence derive the capacity region for the degraded Gaussian channel and feedback channel. Even for the non-degraded Gaussian channel, we will show that this achievable rate region is close to the outer bound on the capacity region.

Other achievable regions can also be derived based on the relay node (user 1) using other relaying schemes to assist user 2, for example, the estimate-and-forward scheme, the amplify-and-forward scheme, or combinations of these schemes. The derivations of these achievable rate regions follow steps that are similar to those used in deriving the achievable rates based on these relaying schemes for the three-terminal relay channel as in [5, 24]. Which scheme results in the largest achievable rate region depends on the particular channel of interest. In general, none of these schemes provide a tight achievable rate region, i.e., the capacity region. Hence, rather than obtaining and comparing the achievable rate regions based on different relaying schemes for a given channel, deriving a relatively tight outer bound on the capacity region for the general channel model is more of interest to us.

For the general discrete memoryless Partially Cooperative RBC, we provide the following outer bound on the capacity region.

**Theorem 2.** *(Outer bound on Capacity Region for Partially Cooperative RBC) The capacity region of the Partially Cooperative RBC is outer bounded by the region with the rate tuples $(R_0, R_1, R_2)$*

*satisfying*

$$\begin{aligned}
&R_0 + R_2 < \min\{I(U, X_1; Y_2), I(U; Y_1, Y_2 \mid X_1)\}, \\
&R_1 < I(X; Y_1, Y_2 \mid U, X_1), \\
&R_0 + R_1 < I(U'; Y_1 | X_1), \\
&R_2 < I(X; Y_1, Y_2 | U', X_1)
\end{aligned} \quad (5)$$

*for some joint distribution $p(u, u', x_1, x)p(y_1, y_2|x, x_1)$ that satisfies two Markov chain conditions: $X_1 \to U \to X$ and $X_1 \to U' \to X$. The auxiliary random variables $U$ and $U'$ are bounded in cardinality by $|\mathcal{U}| \leq |\mathcal{X}| \cdot |\mathcal{X}_1| + 2$ and $|\mathcal{U}'| \leq |\mathcal{X}| \cdot |\mathcal{X}_1| + 2$, respectively.*

*Proof.* See Appendix B. □

**Remark 1.** *The outer bound given in Theorem 2 is tighter than the outer bound based on the general max-flow min-cut theorem [6, Theorem 14.10.1] (the cut-set bound).*

To compare the two outer bounds, we first write the cut-set bound as follows:

$$\begin{aligned}
&R_0 + R_1 < I(X; Y_1 | X_1) \\
&R_0 + R_2 < I(X, X_1; Y_2) \\
&R_0 + R_1 + R_2 < I(X; Y_1, Y_2 | X_1)
\end{aligned} \quad (6)$$

for some joint distribution $p(x, x_1)p(y_1, y_2|x, x_1)$.

We now show that the outer bound given in (5) in Theorem 2 is contained in (tighter than) the cut-set bound given in (6).

In (5), the bound on $R_0 + R_1$ implies

$$\begin{aligned}
R_0 + R_1 &< I(U'; Y_1 | X_1) = H(Y_1 | X_1) - H(Y_1 | U', X_1) \\
&\leq H(Y_1 | X_1) - H(Y_1 | U', X_1, X) \\
&= H(Y_1 | X_1) - H(Y_1 | X_1, X) \\
&= I(X; Y_1 | X_1)
\end{aligned} \quad (7)$$

which is the first cut-set bound given in (6).

In (5), the bound on $R_0 + R_2$ based on the first term in the "min" implies

$$\begin{aligned}
R_0 + R_2 &< I(U, X_1; Y_2) = H(Y_2) - H(Y_2 | U, X_1) \\
&\leq H(Y_2) - H(Y_2 | U, X_1, X) \\
&= H(Y_2) - H(Y_2 | X_1, X) \\
&= I(X, X_1; Y_2)
\end{aligned} \quad (8)$$

which is exactly the second cut-set bound given in (6).

In (5), we combine the bound on $R_0 + R_2$ based on the second term in the "min" with the bound on $R_1$ and have

$$\begin{aligned}
R_0 + R_1 + R_2 &< I(U; Y_1, Y_2 | X_1) + I(X; Y_1, Y_2 | U, X_1) \\
&= I(U, X; Y_1, Y_2 | X_1) \\
&= I(X; Y_1, Y_2 | X_1)
\end{aligned} \quad (9)$$

where we have used the Markov property: $U \to (X_1, X) \to (Y_1, Y_2)$ in the last equality. The bound we have derived in the preceding equation is exactly the third cut-set bound given in (6). Therefore, we conclude that the outer bound given in Theorem 2 is tighter than the cut-set bound.

The inner and outer bounds given in Theorems 1 and 2 may not be tight for the general Partially Cooperative RBC. However, for the degraded Partially Cooperative RBC, which satisfies the condition given in Definition 3, we immediately see that the outer bound (bounds on $R_0 + R_2$ and $R_1$) given in Theorem 2 reduces to the form that matches the inner bound given in Theorem 1. Thus we have the following capacity region for the degraded Partially Cooperative RBC.

**Theorem 3.** *(Capacity Region for Degraded Partially Cooperative RBC) For the degraded Partially Cooperative RBC, which satisfies the condition given in Definition 3, the capacity region is given by the region with the rate tuples $(R_0, R_1, R_2)$ satisfying*

$$\begin{aligned} R_0 + R_2 &< \min \left\{ I(U, X_1; Y_2), I(U; Y_1 \mid X_1) \right\}, \\ R_1 &< I(X; Y_1 \mid U, X_1) \end{aligned} \quad (10)$$

*for some joint distribution $p(x_1)p(u|x_1)p(x|u)p(y_1, y_2|x, x_1)$, where $U$ is bounded in cardinality by $|\mathcal{U}| \leq |\mathcal{X}| \cdot |\mathcal{X}_1| + 2$. Two Markov chains are implied in the joint distribution: $X_1 \to U \to X$ and $U \to (X_1, X) \to (Y_1, Y_2)$.*

*Proof.* The achievability is given by Theorem 1. The converse part can be easily derived from the outer bound for general Partially Cooperative RBCs given in Theorem 2, and a sketch of the proof is given in Appendix C. □

**Remark 2.** *The bounds on $R_0 + R_1$ and on $R_2$ given in Theorem 2 are not useful for the degraded channel. They are still useful for the non-degraded channels as we will demonstrate in the following subsection via the AWGN Partially Cooperative RBC.*

### 2.3 Gaussian Partially Cooperative RBCs

We study two Gaussian Partially Cooperative RBCs, where the outputs at the two users are corrupted by additive white Gaussian noises.

We first define what it means for one Gaussian noise variable to be degraded with respect to another, and then define the two Gaussian Partially Cooperative RBCs with degraded noise terms and independent noise terms, respectively.

**Definition 4.** *The zero mean Gaussian random variable $Z_2$ is (physically) degraded with respect to the zero mean Gaussian random variable $Z_1$ if $Z_2$ can be expressed as $Z_2 = Z_1 + Z'$, where $Z'$ is a zero mean Gaussian random variable that is independent of $Z_1$.*

**Definition 5.** *The D-AWGN Partially Cooperative RBC is a Partially Cooperative RBC with the channel outputs being corrupted by degraded Gaussian noise terms, i.e., the channel outputs at the two users are given by*

$$\begin{aligned} Y_1 &= X + Z_1 \\ Y_2 &= X + X_1 + Z_1 + Z' \end{aligned} \quad (11)$$

*where $Z_1$ and $Z'$ are independent zero mean real Gaussian random variables with variances $N_1$ and $N_2 - N_1$, respectively, where $N_1 < N_2$. The channel input sequences $\{x_n\}$ and $\{x_{1,n}\}$ are subject to* the average power constraints $P$ and $P_1$, respectively, i.e.,

$$\frac{1}{n}\sum_{i=1}^{n} x_i^2 \leq P, \quad \text{and} \quad \frac{1}{n}\sum_{i=1}^{n} x_{1,i}^2 \leq P_1. \tag{12}$$

**Definition 6.** *The AWGN Partially Cooperative RBC is the Partially Cooperative RBC with the channel outputs being corrupted by independent Gaussian noise terms, i.e., the channel outputs at the two users are given by*

$$\begin{aligned} Y_1 &= X + Z_1 \\ Y_2 &= X + X_1 + Z_2 \end{aligned} \tag{13}$$

*where $Z_1$ and $Z_2$ are independent zero mean real Gaussian random variables with variances $N_1$ and $N_2$, respectively, where $N_1 < N_2$. The channel input sequences $\{x_n\}$ and $\{x_{1,n}\}$ are subject to the power constraints given in (12).*

Note that the D-AWGN Partially Cooperative RBC is degraded (satisfies the condition given in Definition 3) due to the degraded Gaussian noise terms at the two outputs. For the D-AWGN Partially Cooperative RBC, we have the following theorem for the capacity region.

**Theorem 4.** *(Capacity Region for D-AWGN Partially Cooperative RBC) The capacity region for the D-AWGN cooperative RBC is given by the region with the rate tuples $(R_0, R_1, R_2)$ satisfying*

$$\begin{aligned} R_0 + R_2 &< \max_{0 \leq \beta \leq 1} \min\left\{ \mathcal{C}\left(\frac{P_1 + \bar{\alpha}P + 2\sqrt{\bar{\beta}\bar{\alpha}PP_1}}{\alpha P + N_2}\right), \mathcal{C}\left(\frac{\beta\bar{\alpha}P}{\alpha P + N_1}\right) \right\}, \\ R_1 &< \mathcal{C}\left(\frac{\alpha P}{N_1}\right) \end{aligned} \tag{14}$$

*for some $\alpha \in [0,1]$, where $\bar{\alpha} := 1 - \alpha$, $\bar{\beta} := 1 - \beta$ and $\mathcal{C}(x) := \frac{1}{2}\log(1+x)$ as defined at the end of Section 1.*

In (14), the parameter $\alpha$ indicates the fraction of source power that is used to transmit information intended for user 1, and $\beta$ is the correlation coefficient between the source and relay signals.

*Proof.* The proof of the achievability follows by evaluating the mutual information terms in Theorem 1 using the following input distributions: $X_1 \sim N(0, P_1)$, $U' \sim N(0, \beta\bar{\alpha}P)$, $X' \sim N(0, \alpha P)$, where $X_1, U', X'$ are independent. Furthermore, we let $U = \sqrt{\frac{\bar{\beta}\bar{\alpha}P}{P_1}}X_1 + U'$ and $X = U + X'$.

The proof of the converse follows directly from the proof of the converse for the D-AWGN Partially Cooperative RBC with feedback (proof given in Appendix E), because the feedback channel provides an outer bound on the capacity region for the original channel without feedback. □

We now study the property of the boundary of the capacity region for the D-AWGN Partially Cooperative RBC. In the following discussion, we let $R_0 = 0$ for convenience. In (14), the optimization over $\beta$ can be evaluated by considering the following two cases.

**Case 1:** If $\frac{P_1}{N_2 - N_1} \geq \frac{P}{N_1}$, then $\beta = 1$ achieves the maximum in (14) for any $\alpha \in [0, 1]$, and the capacity region is defined by the rate pairs $(R_1, R_2)$ that satisfy

$$R_1 < \mathcal{C}\left(\frac{\alpha P}{N_1}\right), \quad R_2 < \mathcal{C}\left(\frac{\bar{\alpha}P}{\alpha P + N_1}\right). \tag{15}$$

Note that in this case $R_1 + R_2 \leq \mathcal{C}\left(\frac{P}{N_1}\right)$, and hence the boundary of the capacity region is a straight line. The capacity region for the D-AWGN Partially Cooperative RBC hence coincides with the capacity region of a broadcast channel where the two users have symmetric channels (both have noise levels $N_1$). This means that if the relay power is large enough, user 2 effectively sees the same level of noise as user 1 due to relaying. Also note that the value $\frac{P(N_2-N_1)}{N_1}$ is a threshold on $P_1$ beyond which the capacity region will not be further enhanced by relaying.

**Case 2:** If $\frac{P_1}{N_2-N_1} < \frac{P}{N_1}$, then define $\alpha_0 := \frac{\frac{P}{N_1} - \frac{P_1}{N_2-N_1}}{\frac{P}{N_1}\frac{P_1}{N_2-N_1} + \frac{P}{N_1}}$. The optimizing $\beta$ will depend on the value of $\alpha$ compared to $\alpha_0$.

(i) If $\alpha \geq \alpha_0$, then $\beta = 1$ achieves the maximum in (14) and again the rate pair given in (15) defines one part of the boundary of the capacity region corresponding to $R_1 \geq \mathcal{C}\left(\frac{\alpha_0 P}{N_1}\right)$. It is clear that this part of the boundary is straight line.

(ii) If $0 \leq \alpha < \alpha_0$, then $\beta^*$ that achieves the maximum satisfies the following equation:

$$\mathcal{C}\left(\frac{P_1 + \bar{\alpha}P + 2\sqrt{\beta\bar{\alpha}PP_1}}{\alpha P + N_2}\right) = \mathcal{C}\left(\frac{\beta\bar{\alpha}P}{\alpha P + N_1}\right). \tag{16}$$

Hence the other part of the boundary of the capacity region corresponding to $0 \leq R_1 \leq \left(\frac{\alpha_0 P}{N_1}\right)$ is defined by the rate pairs $(R_1, R_2)$ that satisfy

$$R_1 < \mathcal{C}\left(\frac{\alpha P}{N_1}\right), \qquad R_2 < \mathcal{C}\left(\frac{\beta^* \bar{\alpha} P}{\alpha P + N_1}\right). \tag{17}$$

We summarize the properties of the boundary of the capacity region in the following proposition.

**Proposition 1.** *If $\frac{P_1}{N_2-N_1} \geq \frac{P}{N_1}$, the boundary of the capacity region for the D-AWGN Partially Cooperative RBC is a straight line defined by (15). If $\frac{P_1}{N_2-N_1} < \frac{P}{N_1}$, the boundary of the capacity region for the D-AWGN Partially Cooperative RBC consists of one straight line segment defined by (15), where $R_1 \geq \mathcal{C}\left(\frac{\alpha_0 P}{N_1}\right)$; and one curved segment defined by (17), where $0 \leq R_1 \leq \left(\frac{\alpha_0 P}{N_1}\right)$.*

We now compare the capacity region for the D-AWGN Partially Cooperative RBC with the capacity region for the Gaussian broadcast channel without user cooperation. The capacity region of the latter channel is given by [6, Chapter 14.6]

$$R_1 < \mathcal{C}\left(\frac{\alpha P}{N_1}\right), \qquad R_2 < \mathcal{C}\left(\frac{\bar{\alpha} P}{\alpha P + N_2}\right) \tag{18}$$

for some $\alpha \in [0,1]$. In Figure 4, we plot this region with the dashed curve as its boundary. We also plot the capacity regions (boundaries with solid lines) for the D-AWGN Partially Cooperative RBC under different relay SNR's $\frac{P_1}{N_2}$. Note that for simplicity, we only plot the region for the case where $R_0 = 0$. It is clear from the figure that the D-AWGN Partially Cooperative RBC has a larger capacity region, and the improvement becomes more significant as $\frac{P_1}{N_2}$ increases. However, as suggested by discussion under Case 1, no further improvement is possible for values of $\frac{P_1}{N_2}$ greater than 14.54 dB. Thus the outer most solid line with $\frac{P_1}{N_2} = 15$dB defines the best capacity region.

We now consider the AWGN Partially Cooperative RBC defined in Definition 6, where the noise terms at the two outputs are independent. This channel does not satisfy the degradedness condition given in Definition 3. An achievable rate region for this channel can be derived, which is exactly the same as the capacity region for the D-AWGN Partially Cooperative RBC.

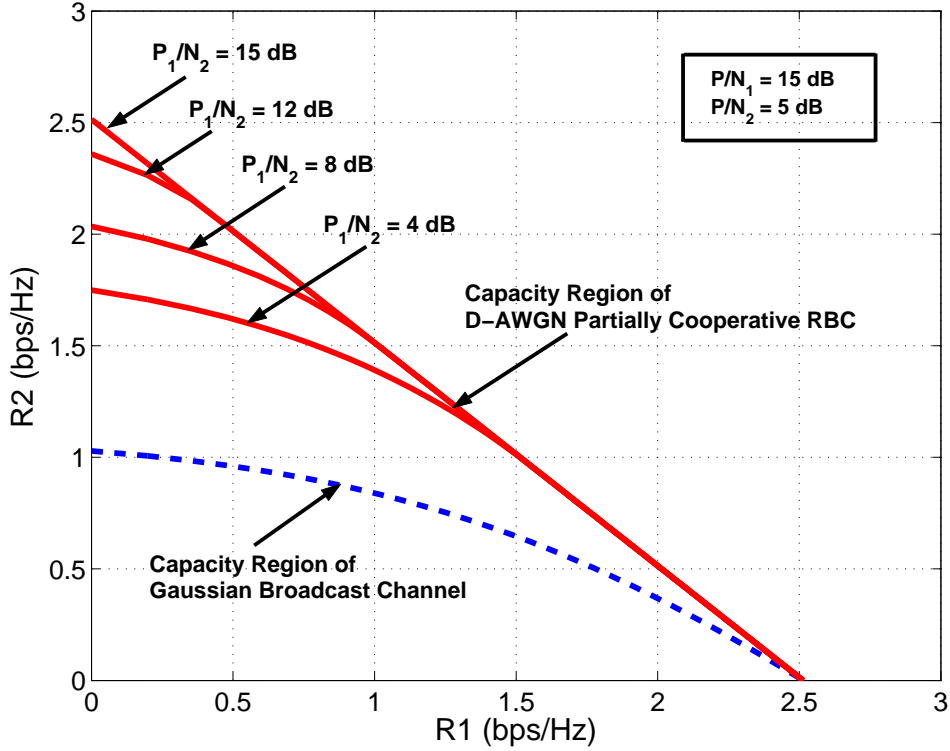

Figure 4: Comparison of the capacity regions for the Gaussian broadcast channel and the D-AWGN Partially Cooperative RBCs.

**Corollary 1.** *(Inner bound on Capacity Region for AWGN Partially Cooperative RBC) An achievable rate region for the AWGN Partially Cooperative RBC is given by the capacity region for the D-AWGN Partially Cooperative RBC given in Theorem 4. Hence the boundary of this achievable rate region has the properties described in Proposition 1.*

The proof follows the steps that are same as in the achievability proof for Theorem 4.

The achievable region given in Corollary 1 may not be a tight inner bound on the capacity region for the AWGN Partially Cooperative RBC. In the following, we further provide an outer bound on the capacity region for this channel.

**Theorem 5.** *(Outer bound on Capacity Region for AWGN Partially Cooperative RBC) The capacity region for the AWGN Partially Cooperative RBC is outer bounded by the rate region with rate tuples $(R_0, R_1, R_2)$ satisfying:*

$$R_0 + R_2 < \min\left\{ \mathcal{C}\left(\frac{P_1 + \bar{\alpha}P + 2\sqrt{\bar{\beta}\bar{\alpha}PP_1}}{\alpha P + N_2}\right), \mathcal{C}\left(\frac{\beta\bar{\alpha}P}{\alpha P + \frac{N_1 N_2}{N_1 + N_2}}\right) \right\},$$
$$R_1 < \mathcal{C}\left(\frac{\alpha P}{\frac{N_1 N_2}{N_1 + N_2}}\right), \quad (19)$$
$$R_0 + R_1 < \mathcal{C}\left(\frac{\alpha P + \beta\bar{\alpha}P}{N_1}\right),$$

*for some $\alpha \in [0,1]$ and $\beta \in [0,1]$, where $\bar{\alpha} := 1 - \alpha$, $\bar{\beta} := 1 - \beta$ and $\mathcal{C}(x) := \frac{1}{2}\log(1+x)$ as defined*

*at the end of Section 1.*

*Proof.* The proof is relegated to Appendix F since part of the proof is similar to the proof of Theorem 7 given in Appendix E. □

In Figure 5, we plot the inner bound on the capacity region (achievable region) of the AWGN Partially Cooperative RBC with dot-dashed line as its boundary and the outer bound on the capacity region with the solid line as its boundary. We plot the region for the case where $R_0 = 0$ for simplicity. It is clear from the figure that the outer and inner bounds are very close. The gap between the two bounds varies with the particular SNRs chosen for the transmission links in the system, and in general is small. In Figure 5, we also plot the capacity region of the original broadcast channel with dashed line as its boundary. It is clear that the AWGN Partially Cooperative RBC has a significantly larger capacity region than the broadcast channel.

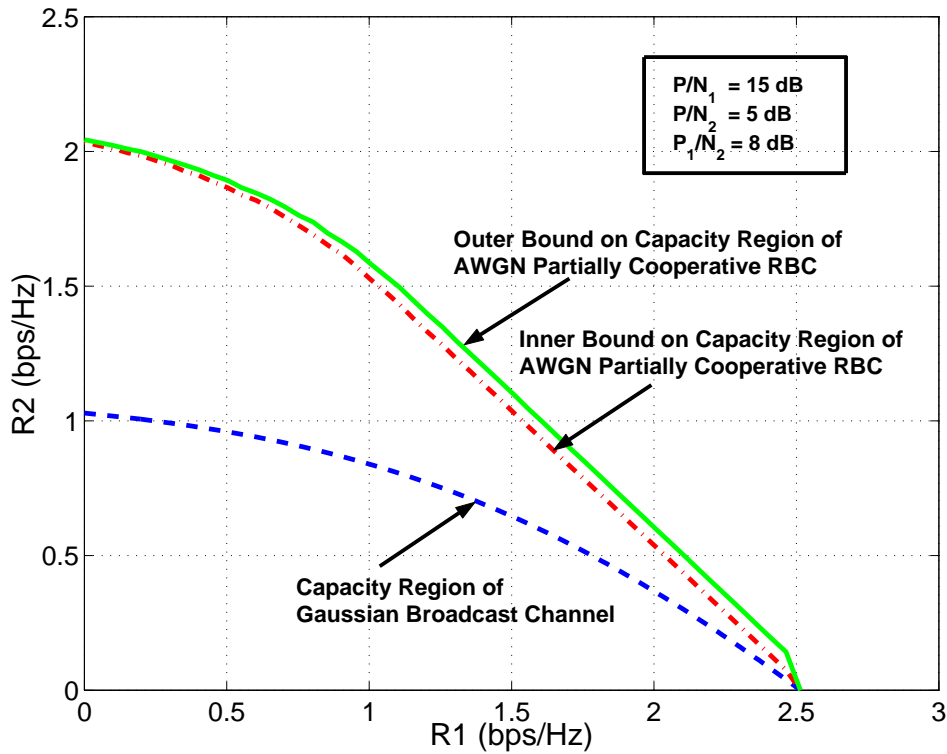

Figure 5: Inner and outer bounds on the capacity region of the AWGN Partially Cooperative RBC.

**Remark 3.** *For the AWGN Partially Cooperative RBC, we have restricted our attention to the channel where $N_1 < N_2$, i.e., the channel from the source to the relay (user 1) is stronger than the channel from the source to user 2. This is a case for which it is reasonable to introduce a relay transmission from user 1 to user 2. Nevertheless, even if the relay (user 1) has a weaker channel from the source than user 2, i.e., $N_1 > N_2$, it can still assist user 2. However, under this condition, the relay needs to use schemes other than the decode-and-forward scheme. For example, the relay can employ the estimate-and-forward scheme to assist user 2, and an achievable rate region based* *on this scheme is given as follows:*

$$\bigcup_{0\leq\alpha\leq 1, 0\leq\eta\leq 1} \left\{ (R_0, R_1, R_2) : \begin{array}{l} R_0 + R_1 < \min\left\{ \mathcal{C}\left(\frac{\bar{\alpha}P}{\alpha P + N_1}\right), \mathcal{C}\left(\frac{\bar{\alpha}P}{\alpha P + \eta P_1 + N_2}\right) \right\}, \\ R_2 < \mathcal{C}\left(\frac{\alpha P}{N_2} + \frac{\alpha \eta P P_1}{\eta P_1 N_1 + \alpha P(N_1 + N_2) + N_1 N_2}\right) \end{array} \right\}. \quad (20)$$

*It is clear that the above achievable region is larger than the capacity region of the original broadcast channel. This achievable rate region can be viewed as a special case of the region given in Theorem 14, with the roles of user 1 and user 2 being switched and the corresponding notations for rates and noise variances also being switched.*

## 3 Partially Cooperative RBCs with Feedback

In this section, we study the Partially Cooperative RBC with feedback, where the outputs at user 2 are provided to user 1 and the outputs at both users 1 and 2 are provided to the source all through perfect feedback links (see Figure 6). Note that this definition for feedback channel follows the definition for the relay channel with feedback [5, Section V]. We will show that feedback in general may improve the capacity region for the Partially Cooperative RBCs, but does not affect the capacity region for the degraded Partially Cooperative RBCs.

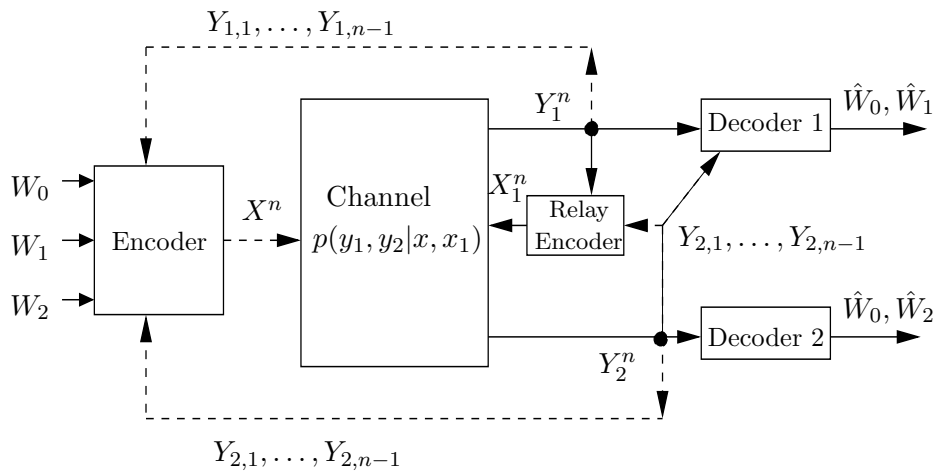

Figure 6: Partially Cooperative RBC with feedback

For a distribution on the message set $p(w_0, w_1, w_2)$, the following joint distribution is induced for a Partially Cooperative RBC with feedback:

$$p(w_0, w_1, w_2, x^n, x_1^n, y_1^n, y_2^n)$$
$$= p(w_0, w_1, w_2) \prod_{i=1}^{n} p(x_i|w_0, w_1, w_2, y_1^{i-1}, y_2^{i-1}) p(x_{1,i}|y_1^{i-1}, y_2^{i-1}) p(y_{1,i}, y_{2,i}|x_i, x_{1,i}) \quad (21)$$

### 3.1 Discrete Memoryless Partially Cooperative RBCs with Feedback

From the definition for the degraded Partially Cooperative RBC given in Definition 3, it is clear that the Partially Cooperative RBC with feedback is degraded. For this channel, we have the following capacity theorem.

**Theorem 6.** *(Capacity Region for Partially Cooperative RBC with Feedback) The capacity region of the Partially Cooperative RBC with Feedback is given by the rate region with the rate tuples $(R_0, R_1, R_2)$ satisfying*

$$\begin{aligned} R_0 + R_2 &< \min \left\{ I(U, X_1; Y_2), I(U; Y_1, Y_2 \mid X_1) \right\}, \\ R_1 &< I(X; Y_1, Y_2 \mid U, X_1) \end{aligned} \tag{22}$$

*for some joint distribution $p(x_1)p(u|x_1)p(x|u)p(y_1, y_2|x, x_1)$, where $U$ is bounded in cardinality by $|\mathcal{U}| \leq |\mathcal{X}| \cdot |\mathcal{X}_1| + 2$.*

*Proof.* Theorem 1 provides the achievability with $(Y_1, Y_2)$ replacing $Y_1$ as the output at user 1. The proof of the converse follows steps that are similar to those in the proof for Theorem 2, and is hence only outlined in Appendix D. □

**Remark 4.** *From the above achievability proof, it is clear that in the capacity region achieving scheme, the source does not exploit the feedback information from the two users. However, the relay (user 1) makes use of the feedback information from user 2 to improve on its decoding and relaying.*

It is clear that if the original Partially Cooperative RBC is degraded, then the capacity region given in Theorem 6 is the same as the capacity region for the original channel without feedback.

**Corollary 2.** *Feedback does not increase the capacity region for the degraded Partially Cooperative RBCs.*

This result is intuitive. Since the original channel is degraded, the output $y_2$ at user 2 does not provide user 1 with more information other than the information already contained in the output $y_1$ at user 1. Hence feedback of $y_2$ to user 1 does not help. The reason that feedback of $y_1$ and $y_2$ to the source does not help follows from the result in [10] that feedback does not increase the capacity for the physically degraded broadcast channel.

## 3.2 Gaussian Partially Cooperative RBCs with Feedback

In this section, we consider two Gaussian feedback channels: the D-AWGN Partially Cooperative RBC with feedback and the AWGN Partially Cooperative RBC with feedback. We study how feedback affects the capacity regions of these Gaussian channels.

For the D-AWGN Partially Cooperative RBC with feedback, we have the following theorem, which is consistent with the result given in Corollary 2 for the discrete memoryless channels.

**Theorem 7.** *(Capacity Region for D-AWGN Partially Cooperative RBC with Feedback) Feedback does not increase the capacity region for the D-AWGN Partially Cooperative RBC, i.e., the capacity region for the D-AWGN Partially Cooperative RBC with feedback is the same as the capacity region for the D-AWGN Partially Cooperative RBC without feedback given in Theorem 4.*

*Proof.* The achievability proof is same as that for the D-AWGN Partially Cooperative RBC. The proof for the converse is provided in Appendix E. □

We note that in the definition for the D-AWGN Partially Cooperative RBC, the output $y_1$ at user 1 is not affected by the input signal $x_1$ transmitted by user 1 to user 2. In practice, $x_1$ may cause interference to $y_1$. We hence define the following *Self-Interfered D-AWGN Partially Cooperative RBC* model:

$$Y_1 = X + aX_1 + Z_1$$
$$Y_2 = X + X_1 + Z_1 + Z' \quad (23)$$

where $a$ is a real constant number indicating how strong the interference is, and $Z_1$ and $Z'$ are independent zero mean real Gaussian random variables with variances $N_1$ and $N_2 - N_1$, respectively.

The Self-Interfered D-AWGN Partially Cooperative RBC is degraded. It is also easy to check that the self-interference does not affect the capacity region of the D-AWGN Partially Cooperative RBC with/without feedback. This result is summarized in the following corollary, and it will be useful in proving the capacity region for the AWGN Partially Cooperative RBC with feedback given in Theorem 8.

**Corollary 3.** *The capacity region for the Self-Interfered D-AWGN Partially Cooperative RBC with/without feedback is the same as the capacity region for the D-AWGN Partially Cooperative RBC with/without feedback, and is given in Theorem 4.*

We now consider the second Gaussian feedback channel: the AWGN Partially Cooperative RBC with feedback. For this channel, we have the following capacity theorem.

**Theorem 8.** *(Capacity Region for AWGN Partially Cooperative RBC with Feedback) The capacity region for the AWGN Partially Cooperative RBC with feedback is given by*

$$R_0 + R_2 < \max_{0 \leq \beta \leq 1} \min \left\{ \mathcal{C}\left(\frac{P_1 + \bar{\alpha}P + 2\sqrt{\bar{\beta}\bar{\alpha}PP_1}}{\alpha P + N_2}\right), \mathcal{C}\left(\frac{\beta\bar{\alpha}P}{\alpha P + \frac{N_1 N_2}{N_1 + N_2}}\right) \right\},$$
$$R_1 < \mathcal{C}\left(\frac{\alpha P}{\frac{N_1 N_2}{N_1 + N_2}}\right) \quad (24)$$

*for some $\alpha \in [0, 1]$.*

*Proof.* The idea of the proof is to follow the argument in [36, Chapter 3.2.2] to change the AWGN Partially Cooperative RBC with feedback to an equivalent D-AWGN Partially Cooperative RBC with feedback, with $N_1$ being replaced by $\frac{N_1 N_2}{N_1 + N_2}$.

We first define

$$S := \frac{N_1 Y_2 + N_2 Y_1}{N_1 + N_2} \quad (25)$$

and note that the mapping from $(Y_1, Y_2)$ to $(S, Y_2)$ is one-to-one. Hence the channel with the outputs being $(Y_1, Y_2)$ and $Y_2$ is equivalent to the channel with the outputs being $(S, Y_2)$ and $Y_2$.

We now want to show that given $(S, X_1)$, $Y_2$ is independent of $X$, i.e., $Y_2$ is a degraded version of $S$. We express $S$ in the following form:

$$S = X + \frac{N_1}{N_1 + N_2}X_1 + \hat{Z}_1 \quad (26)$$

where $\hat{Z}_1 := \frac{N_1 Z_2 + N_2 Z_1}{N_1 + N_2}$. We can express $Y_2$ as

$$Y_2 = S + \frac{N_2}{N_1 + N_2}X_1 + \hat{Z} \quad (27)$$

where $\hat{Z} := \frac{N_2}{N_1+N_2}(Z_2 - Z_1)$.

It is clear that $\hat{Z}$ is independent of $X$ and $X_1$, and it is easy to check that $\hat{Z}$ is independent of $\hat{Z}_1$. Hence $\hat{Z}$ is independent of $S$. Therefore, given $(S, X_1)$, $Y_2$ is independent of $X$.

Now, for the equivalent channel with outputs being $(S, Y_2)$ and $Y_2$, we have

$$\begin{aligned} S &= X + \frac{N_1}{N_1 + N_2}X_1 + \hat{Z}_1 \\ Y_2 &= X + X_1 + \hat{Z}_1 + \hat{Z} \end{aligned} \quad (28)$$

where $\hat{Z}_1$ and $\hat{Z}$ are independent zero mean real Gaussian random variables with variances $\frac{N_1 N_2}{N_1+N_2}$ and $\frac{N_2^2}{N_2+N_1}$, respectively.

This is a Self-Interfered D-AWGN Partially Cooperative RBC with feedback, and hence Corollary 3 can be applied to obtain the capacity region. □

**Corollary 4.** *Feedback enlarges the capacity region for the AWGN Partially Cooperative RBC.*

This corollary can be shown by comparing the capacity region of the AWGN Partially Cooperative RBC with feedback given in Theorem 8 with the outer bound on the capacity region of the AWGN Partially Cooperative RBC given in Theorem 5. This result is reasonable because for the AWGN Partially Cooperative RBC, the output $y_2$ at user 2 is not a degraded version of the output $y_1$ at user 1, and hence feedback of $y_2$ to user 1 provides further information and results in an enlargement of the capacity region.

## 3.3 Comparison with Results on Broadcast Channels with Feedback

At this point, it is instructive to compare the results on the capacity regions for broadcast channels with feedback (see [4] for a review) with our results on the capacity regions of the Partially Cooperative RBCs with feedback.

We have obtained the capacity region of the general discrete memoryless Partially Cooperative RBC with feedback in Theorem 6. However, the capacity region is still not known for the general discrete memoryless broadcast channel with feedback. We have shown that feedback does not enlarge the capacity region for the degraded Partially Cooperative RBC and its Gaussian example in Corollary 2 and Theorem 7. This result is consistent with the result obtained in [10, 11] that feedback does not enlarge the capacity region for the physically degraded broadcast channel.

We have obtained the capacity region of the AWGN Partially Cooperative RBC with feedback in Theorem 8. However, the capacity region of the AWGN broadcast channel with feedback is still not known. We have shown that feedback enlarges the capacity region for the AWGN Partially Cooperative RBC. This is consistent with the result in [37] that feedback enlarges the capacity region for the AWGN broadcast channel. However, the reasons for the enlargement are different for the two channels. For the AWGN Partially Cooperative RBC with feedback, the source does not make use of the feedback information. It is user 1 that utilizes the feedback information from user 2 to improve on its decoding and relaying. Such a strategy achieves the capacity region. Whereas for the broadcast channel with feedback, the source needs to make use of the feedback information to improve the encoding. In [37], the authors provided an example encoding scheme for the source to exploit feedback information to improve the capacity region. However, the optimal encoding scheme that achieves the capacity region for the AWGN broadcast channel with feedback is still not known.

We finally note that the structure of feedback in Partially Cooperative RBCs is different from that in broadcast channels. In Partially Cooperative RBCs with feedback, the output at user 2 is also fed back to user 1, but this feedback is not available in broadcast channels with feedback. Hence obtaining the capacity region for the Partially Cooperative RBCs with feedback does not necessarily imply obtaining the capacity region for the broadcast channel with feedback.

# 4 Fully Cooperative Relay Broadcast Channels

In the previous sections, we studied the Partially Cooperative RBC where one user (usually the user with "better channel" from the source) in the broadcast system helps the other user by sending relay signals. In this case, we have seen that the Partially Cooperative RBC has a larger capacity region than the original broadcast channel due to user cooperation. It is then natural to explore whether the capacity region can be further enlarged if we allow both users to help each other by sending cooperative signals through relay links.

In this section, we study the Fully Cooperative RBC, where not only user 1 serves as a relay to help user 2, but user 2 serves as a relay to assist user 1 as well. We will first describe the channel model, and then present our main results. We further illustrate the results on achievable rate/capacity regions via Gaussian examples, and compare these regions with those of the Partially Cooperative RBC.

## 4.1 System Model

We use $x$ to denote the source input, $x_1$ and $x_2$ to denote the relay inputs from users 1 and 2, respectively, and $y_1$ and $y_2$ to denote the outputs at users 1 and 2, respectively.

**Definition 7.** *A Fully Cooperative RBC consists of a channel input alphabet $\mathcal{X}$, two relay input alphabets $\mathcal{X}_1$ and $\mathcal{X}_2$, two channel output alphabets $\mathcal{Y}_1$ and $\mathcal{Y}_2$, and a probability transition function $p(y_1, y_2 | x, x_1, x_2)$ (see Figure 7).*

**Definition 8.** *A Fully Cooperative RBC is degraded if it either satisfies the condition*

$$p(y_1, y_2|x, x_1, x_2) = p(y_1|x, x_1, x_2)p(y_2|y_1, x_1, x_2), \qquad (29)$$

*i.e., $y_2$ is independent of $x$, conditioned on $y_1$, $x_1$ and $x_2$; or satisfies the condition*

$$p(y_1, y_2|x, x_1, x_2) = p(y_2|x, x_1, x_2)p(y_1|y_2, x_1, x_2), \qquad (30)$$

*i.e., $y_1$ is independent of $x$, conditioned on $y_2$, $x_1$ and $x_2$.*

Without loss of generality, in this paper we only considers the degraded channel that satisfies the first condition.

The definition for a $(2^{nR_0}, 2^{nR_1}, 2^{nR_2}, n)$ code for the Fully Cooperative RBC is similar to that for the Partially Cooperative RBC, except that it includes another set of relay functions $\{g_i\}_{i=1}^n$ such that

$$x_{2,i} = g_i(y_{2,1}, \ldots, y_{2,i-1}), \qquad 1 \leq i \leq n. \qquad (31)$$

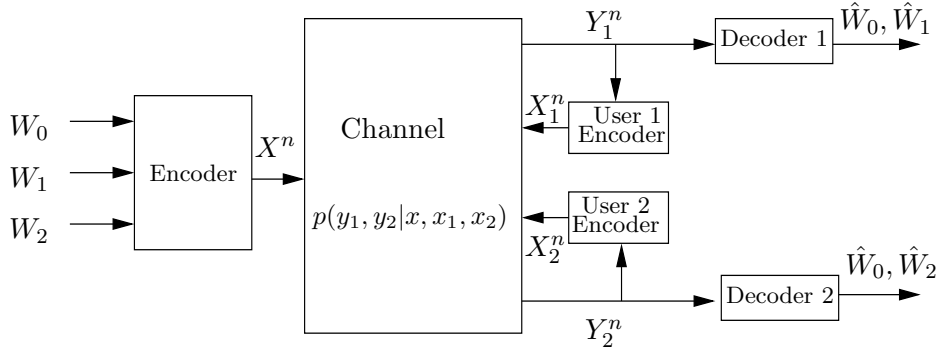

Figure 7: Fully Cooperative Relay Broadcast Channel

## 4.2 Discrete Memoryless Fully Cooperative RBCs

To derive an achievable rate region for the Fully Cooperative RBC, we first need to choose relaying schemes for user 1 and user 2 to assist each other. A simple choice is one where one of the users employs the decode-and-forward scheme, and the other user always sends a single codeword (that may vary according to the target rate tuple) which results in the best achievable rate region. We obtain the following achievable rate region which is the union of two achievable rate regions derived by switching these two relaying schemes for the two users.

**Theorem 9.** *(Inner bound on Capacity Region for Fully Cooperative RBC) An achievable rate region for the Fully Cooperative RBC is the convex hull of the union of the following two rate regions $\mathcal{R}_1$ and $\mathcal{R}_2$:*

$$\mathcal{R}_1 = \bigcup_{p(u,x_1,x_2,x)} \left\{ (R_0, R_1, R_2) : \begin{array}{l} R_0 + R_2 < \min\{I(U, X_1; Y_2|X_2), I(U; Y_1|X_1, X_2)\}, \\ R_1 < I(X; Y_1 \mid U, X_1, X_2) \end{array} \right\} \quad (32)$$

$$\mathcal{R}_2 = \bigcup_{p(u',x_1,x_2,x)} \left\{ (R_0, R_1, R_2) : \begin{array}{l} R_0 + R_1 < \min\{I(U', X_2; Y_1|X_1), I(U'; Y_2|X_1, X_2)\}, \\ R_2 < I(X; Y_2 \mid U', X_1, X_2) \end{array} \right\} \quad (33)$$

*where $U$ and $U'$ are bounded in cardinality by $|\mathcal{U}| \leq |\mathcal{X}| \cdot |\mathcal{X}_1| \cdot |\mathcal{X}_2| + 2$ and $|\mathcal{U}'| \leq |\mathcal{X}| \cdot |\mathcal{X}_1| \cdot |\mathcal{X}_2| + 2$, respectively.*

The proof is similar to the proof for Theorem 1 and is omitted.

The achievable rate region given in Theorem 9 will be shown to be tight for the degraded Fully Cooperative RBC in Theorem 12. However, this achievable rate region may not be tight for a general Fully Cooperative RBC. For example, for the AWGN Fully Cooperative RBC which we will consider later, this achievable rate region is not tight, and the relay node that sends a single codeword does not help at all. This relay needs to employ a better relaying scheme, for example, the estimate-and-forward scheme [5, Theorem 6], to be of help. Hence, motivated by the AWGN Fully Cooperative RBC, we provide the following achievable rate region which is based on the scheme where one user in the system employs the decode-and-forward scheme and the other user in the system employs the estimate-and-forward scheme to relay information.

**Theorem 10.** *(Inner bound on Capacity Region for Fully Cooperative RBC) An achievable rate region for the Fully Cooperative RBC is the convex hull of the union of the following two rate regions*

$\mathcal{R}_1$ and $\mathcal{R}_2$:

$$\mathcal{R}_1 = \bigcup_{\substack{p(x_2)p(u,x_1,x) \\ p(y_1,y_2|x,x_1,x_2) \\ p(\hat{y}_2|y_2,x_1,x_2,u)}} \left\{ (R_0, R_1, R_2) : \begin{array}{l} R_0 + R_2 < \min\{I(U,X_1;Y_2|X_2), I(U;Y_1|X_1)\}, \\ R_1 < I(X;\hat{Y}_2, Y_1|U, X_1, X_2) \\ \texttt{subject to:} \\ I(X_2;Y_1|U,X_1) \geq I(\hat{Y}_2;Y_2|Y_1, U, X_1, X_2) \end{array} \right\}$$
(34)

$$\mathcal{R}_2 = \bigcup_{\substack{p(x_1)p(u',x_2,x) \\ p(y_1,y_2|x,x_1,x_2) \\ p(\hat{y}_1|y_1,x_1,x_2,u')}} \left\{ (R_0, R_1, R_2) : \begin{array}{l} R_0 + R_1 < \min\{I(U', X_2;Y_1|X_1), I(U';Y_2|X_2)\}, \\ R_2 < I(X;\hat{Y}_1, Y_2|U', X_1, X_2) \\ \texttt{subject to:} \\ I(X_1;Y_2|U',X_2) \geq I(\hat{Y}_1;Y_1|Y_2, U', X_1, X_2) \end{array} \right\}$$
(35)

*Proof.* See Appendix G for an outline of the proof. □

The achievable rate region given in Theorem 10 serves as an example to demonstrate that the Fully Cooperative RBC can achieve larger rate region than the Partially Cooperative RBC due to an additional relay link. This will be clear when we apply Theorem 10 to the AWGN Fully Cooperative RBC in the next subsection.

There are other relaying schemes that the system can choose for the two users to assist each other, and each of these relaying schemes results in an achievable rate region. In general, these achievable regions are not tight. As we have remarked for the Partially Cooperative RBC, we will focus on deriving an outer bound on the capacity region instead of providing all the achievable regions that we can think of.

In the following theorem, we provide an outer bound on the capacity region for the Fully Cooperative RBC. Note that this outer bound is tighter than the outer bound given by the cut-set bound.

**Theorem 11.** *(Outer bound on Capacity Region for Fully Cooperative RBC) The capacity region for the Fully Cooperative RBC is outer bounded by the rate region with the rate tuples $(R_0, R_1, R_2)$ satisfying:*

$$\begin{aligned} R_0 + R_2 &< \min\{I(U, X_1; Y_2|X_2), I(U; Y_1, Y_2|X_1, X_2)\}, \\ R_1 &< I(X; Y_1, Y_2|U, X_1, X_2), \\ R_0 + R_1 &< \min\{I(U', X_2; Y_1|X_1), I(U'; Y_1, Y_2|X_1, X_2)\}, \\ R_2 &< I(X; Y_1, Y_2|U', X_1, X_2) \end{aligned}$$
(36)

*for the joint distribution $p(u, u', x_1, x_2, x)p(y_1, y_2|x_1, x_2, x)$ that satisfies the Markov chains: $(X_1, X_2) \to U \to X$ and $(X_1, X_2) \to U' \to X$. Furthermore, the auxiliary random variables $U$ and $U'$ are bounded in cardinality by $|\mathcal{U}| \leq |\mathcal{X}| \cdot |\mathcal{X}_1| \cdot |\mathcal{X}_2| + 2$ and $|\mathcal{U}'| \leq |\mathcal{X}| \cdot |\mathcal{X}_1| \cdot |\mathcal{X}_2| + 2$, respectively.*

The proof for Theorem 11 is similar to the proof for Theorem 2, and hence is omitted.

For the special case of the degraded Fully Cooperative RBC, we have the following capacity theorem.

**Theorem 12.** *(Capacity Region for Degraded Fully Cooperative RBC) The capacity region for the degraded Fully Cooperative RBC that satisfies the condition* (29) *is given by*

$$\bigcup_{p(x_1,x_2,u)p(x|u)} \left\{ (R_0, R_1, R_2) : \begin{array}{l} R_0 + R_2 < \min\left\{I(U, X_1; Y_2|X_2), I(U; Y_1|X_1, X_2)\right\}, \\ R_1 < I(X; Y_1 \mid U, X_1, X_2) \end{array} \right\} \quad (37)$$

*where $U$ is bounded in cardinality by $|\mathcal{U}| \leq |\mathcal{X}| \cdot |\mathcal{X}_1| \cdot |\mathcal{X}_2| + 2$.*

*Proof.* The achievability is given by Theorem 9. The converse proof follows by applying the degraded condition (29) to the first two bounds given in Theorem 11. □

Note that for the degraded Fully Cooperative RBC, the output at user 2 is a degraded version of the output at user 1, and it does not receive any more information than user 1. Hence user 2 does not need to relay any information for user 1, and all it does is to send a single codeword (may vary according to the target rate tuple) which results in the best achievable rate region. The converse for Theorem 12 shows that this scheme is optimal.

## 4.3 Gaussian Fully Cooperative RBCs

As for the Partially Cooperative RBC, we study two Gaussian channels for the Fully Cooperative RBC: the D-AWGN and the AWGN Fully Cooperative RBCs.

For the D-AWGN Fully Cooperative RBC, the channel outputs at the two users are given by

$$\begin{aligned} Y_1 &= X + X_2 + Z_1 \\ Y_2 &= X + X_1 + Z_1 + Z' \end{aligned} \quad (38)$$

where $Z_1$ and $Z'$ are independent zero mean real Gaussian random variables with variances $N_1$ and $N_2 - N_1$, respectively, where $N_1 < N_2$. The channel input sequences $\{x_n\}$, $\{x_{1,n}\}$ and $\{x_{2,n}\}$ are subject to the average power constraints $P$, $P_1$ and $P_2$, respectively, i.e.,

$$\frac{1}{n}\sum_{i=1}^{n} x_i^2 \leq P, \qquad \frac{1}{n}\sum_{i=1}^{n} x_{1,i}^2 \leq P_1, \qquad \text{and} \qquad \frac{1}{n}\sum_{i=1}^{n} x_{2,i}^2 \leq P_2. \quad (39)$$

Note that the D-AWGN Fully Cooperative RBC satisfies condition (29), and is hence degraded. The following theorem provides the capacity region for the D-AWGN Fully Cooperative RBC.

**Theorem 13.** *(Capacity Region for D-AWGN Fully Cooperative RBC) The capacity region for the D-AWGN Fully Cooperative RBC is the same as the capacity region for the D-AWGN Partially Cooperative RBC given in Theorem 4.*

*Proof.* The proof of the achievability is straightforward based on the proof of the achievability for the Partially Cooperative RBC with user 2 being silent. The proof of the converse follows similarly as the proof of the converse for Theorem 7. □

**Remark 5.** *The relay link from user 2 to user 1 does not help to enlarge the capacity region for the D-AWGN Fully Cooperative RBC.*

The intuition behind Theorem 13 is as follows. For the D-AWGN Fully Cooperative RBC, user 2 receives a degraded version of the output at user 1, and hence can not provide further information for user 1 other than the information that user 1 already knows. This result is also consistent with Theorem 7 that feedback does not increase the capacity region for the D-AWGN Partially Cooperative RBC, i.e., perfectly providing the output at user 2 to user 1 cannot enlarge the capacity region for the D-AWGN Partially Cooperative RBC. It is then reasonable that sending information based on the output at user 2 to user 1 through a noisy channel cannot enlarge the capacity region for the D-AWGN Partially Cooperative RBC. This is exactly what Theorem 13 concludes.

However, Theorem 13 is not necessarily true for the discrete memoryless degraded Fully Cooperative RBC. First of all, it is clear that the Fully Cooperative RBC achieves at least the capacity region for the corresponding Partially Cooperative RBC with user 2 "being silent" (sending a fixed alphabet symbol). We now explore whether the degraded Fully Cooperative RBC can achieve a better rate region. Although for the discrete memoryless degraded Fully Cooperative RBC, user 2 still cannot provide further information to user 1 through the relay link, it can affect the channel by sending a predetermined codeword (this is not the case for the D-AWGN Fully Cooperative RBC, because the channel is determined by the Gaussian noises and is independent of the input). Hence user 2 can send a single codeword through the relay link to result in the best achievable region. Therefore, the relay link from user 2 to user 1 can potentially assist in enlarging the capacity region for the discrete memoryless Partially Cooperative RBC.

We now consider the AWGN Fully Cooperative RBC where the Gaussian noise terms at the two receivers are independent. For this channel, the outputs at the two users are given by

$$\begin{aligned} Y_1 &= X + X_2 + Z_1 \\ Y_2 &= X + X_1 + Z_2 \end{aligned} \tag{40}$$

where $Z_1$ and $Z_2$ are independent zero mean real Gaussian random variables with variances $N_1$ and $N_2$, respectively, where $N_1 < N_2$. The channel input sequences $\{x_n\}$, $\{x_{1,n}\}$ and $\{x_{2,n}\}$ are subject to the power constraints given in (39).

It is clear that the AWGN Fully Cooperative RBC can achieve the rate region of the AWGN Partially Cooperative RBC given in Corollary 1 by using the same coding scheme and keeping user 2 silent. Can it achieve a better rate region? Since the outputs at user 1 and user 2 are corrupted by stochastically independent noise terms, user 2 indeed receives some additional information other than the information received at user 1. Hence potentially user 2 can assist in enlarging the rate for user 1 by sending this additional information to user 1 through the relay link. Although user 2 cannot decode this additional information (receiver noise level at user 2 is higher than that at user 1), it can first compress this information and then forward it to user 1, i.e., user 2 can employ the estimate-and-forward relaying scheme.

We provide an achievable rate region for the AWGN Fully Cooperative RBC based on user 1 employing the decode-and-forward relaying scheme and user 2 employing the estimate-and-forward relaying scheme in the following theorem.

**Theorem 14.** *(Inner bound for AWGN Fully Cooperative RBC) An achievable rate region for the AWGN Fully Cooperative RBC is the convex hull of the rate tuples $(R_0, R_1, R_2)$ that satisfy*

$$\begin{aligned} R_1 &< \mathcal{C}\left(\frac{\alpha P}{N_1} + \frac{\alpha \eta P P_2}{\eta P_2 N_2 + \alpha P(N_1 + N_2) + N_1 N_2}\right) \\ R_0 + R_2 &< \max_{0 \le \beta \le 1} \min\left\{\mathcal{C}\left(\frac{\bar{\alpha}P + P_1 + 2\sqrt{\beta\bar{\alpha}PP_1}}{\alpha P + N_2}\right), \mathcal{C}\left(\frac{\beta\bar{\alpha}P}{\alpha P + \eta P_2 + N_1}\right)\right\} \end{aligned} \tag{41}$$

*for some* $\alpha \in [0, 1]$ *and* $\eta \in [0, 1]$.

*Proof.* See Appendix H. □

Note that the parameter $\eta$ in (41) is the fraction of the relay power at user 2 being used for relaying transmission. This parameter can be used to tradeoff between the rates $R_2$ and $R_1$. Enlarging $\eta$ sacrifices $R_2$ to improve $R_1$. This is because larger $\eta$ causes more interference to user 1 in the decoding of the information for user 2, and hence user 1 becomes less helpful for user 2. On the other hand, user 1 gets more benefit from user 2 due to the larger relaying power.

**Remark 6.** *Theorem 14 shows that the AWGN Fully Cooperative RBC indeed achieves a larger rate region than the AWGN Partially Cooperative RBC. This is because the relay link from user 2 to user 1 assists in enlarging the rate region. Theorem 14 also shows that the AWGN Fully Cooperative RBC achieves a larger rate region than the D-AWGN Fully Cooperative RBC. These two facts will be demonstrated by the numerical results at the end of this section.*

We further provide an outer bound on the capacity region for the AWGN Fully Cooperative RBC.

**Theorem 15.** *(Outer bound on Capacity Region for AWGN Fully Cooperative RBC) The capacity region for the AWGN Fully Cooperative RBC is outer bounded by the rate region with rate tuples $(R_0, R_1, R_2)$ satisfying*

$$
\begin{aligned}
R_0 + R_2 &< \min\left\{\mathcal{C}\left(\frac{P_1 + \bar{\alpha}P + 2\sqrt{\beta\bar{\alpha}PP_1}}{\alpha P + N_2}\right), \mathcal{C}\left(\frac{\beta\bar{\alpha}P}{\alpha P + \frac{N_1 N_2}{N_1 + N_2}}\right)\right\}, \\
R_1 &< \mathcal{C}\left(\frac{\alpha P}{\frac{N_1 N_2}{N_1 + N_2}}\right), \\
R_0 + R_1 &< \min\left\{\mathcal{C}\left(\frac{P + P_2 + 2\sqrt{\bar{\beta}\bar{\alpha}PP_2}}{N_1}\right), \mathcal{C}\left(\frac{\bar{\gamma}(\alpha P + \beta\bar{\alpha}P)}{\gamma(\alpha P + \beta\bar{\alpha}P) + \frac{N_1 N_2}{N_1 + N_2}}\right)\right\}, \\
R_2 &< \mathcal{C}\left(\frac{\gamma(\alpha P + \beta\bar{\alpha}P)}{\frac{N_1 N_2}{N_1 + N_2}}\right)
\end{aligned}
\tag{42}
$$

*for some* $\alpha, \beta, \gamma \in [0, 1]$.

*Proof.* See Appendix I. □

We now compare the achievable rate region for the AWGN Fully Cooperative RBC with the capacity region for the D-AWGN Fully/Partially Cooperative RBC numerically. In Figure 8, we plot the inner bound (boundary with solid line) and outer bound (boundary with circled line) on the capacity region for the AWGN Fully Cooperative RBC, the capacity region (boundary with dot-dashed line) for the D-AWGN Fully/Partially Cooperative RBC, and the capacity region (boundary with dashed line) of the original Gaussian broadcast channel without relay links. Since the capacity region of the AWGN Fully Cooperative RBC lies between its inner and outer bounds, it is clear from the figure that this capacity region is larger than the capacity region of the D-AWGN Fully/Partially Cooperative RBC.

In Figure 8, we also plot the achievable region for the AWGN Partially Cooperative RBC (boundary also with dot-dashed line). It is clear from the figure that the achievable region for the AWGN

Fully Cooperative RBC is larger than the achievable region for the AWGN Partially Cooperative RBC. Furthermore, for the AWGN Fully Cooperative RBC, the maximum rates of both users 1 and 2 are improved relative to the original Gaussian broadcast channel. However, for the AWGN Partially Cooperative RBC, the maximum rate of only user 2 is improved. This is because user 2 in the AWGN Fully Cooperative RBC helps user 1 through a relay link, which is not allowed for the AWGN Partially Cooperative RBC.

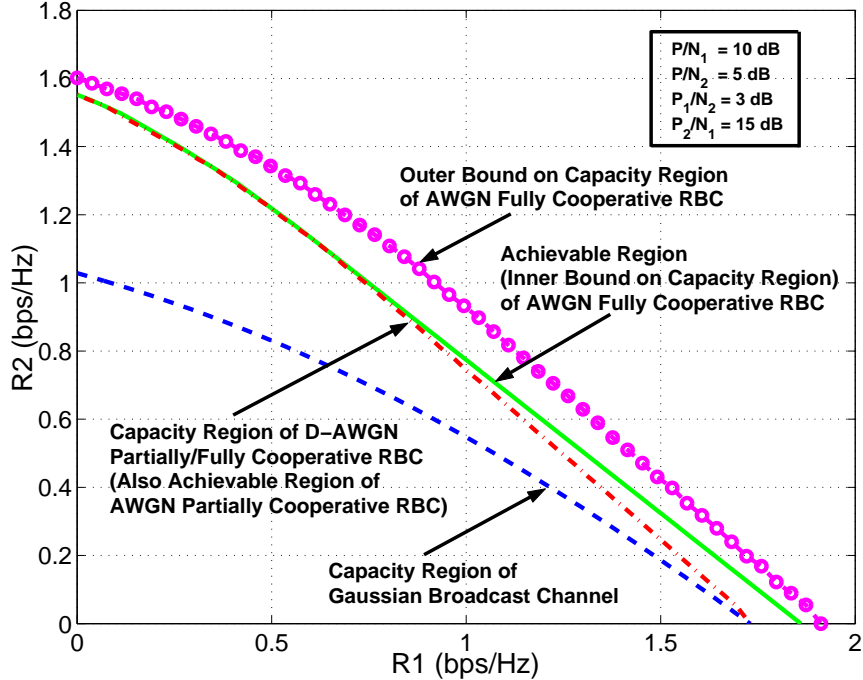

Figure 8: Comparison of rate regions for Gaussian RBCs

## 5 Fully Cooperative RBCs with Feedback

In this section, we study the Fully Cooperative RBC with feedback, where the outputs at user 2 are provided to user 1 and the outputs at both users 1 and 2 are provided to the source all through perfect feedback links (see Figure 9). We study how feedback affects the Fully Cooperative RBC.

For a distribution on the message set $p(w_0, w_1, w_2)$, the following joint distribution is induced for a Fully Cooperative RBC with feedback

$$p(w_0, w_1, w_2, x^n, x_1^n, x_2^n, y_1^n, y_2^n)$$
$$= p(w_0, w_1, w_2) \prod_{i=1}^{n} p(x_i|w_0, w_1, w_2, y_1^{i-1}, y_2^{i-1}) p(x_{1,i}|y_1^{i-1}, y_2^{i-1}) \qquad (43)$$
$$\cdot p(x_{2,i}|y_2^{i-1}) p(y_{1,i}, y_{2,i}|x_i, x_{1,i}, x_{2,i})$$

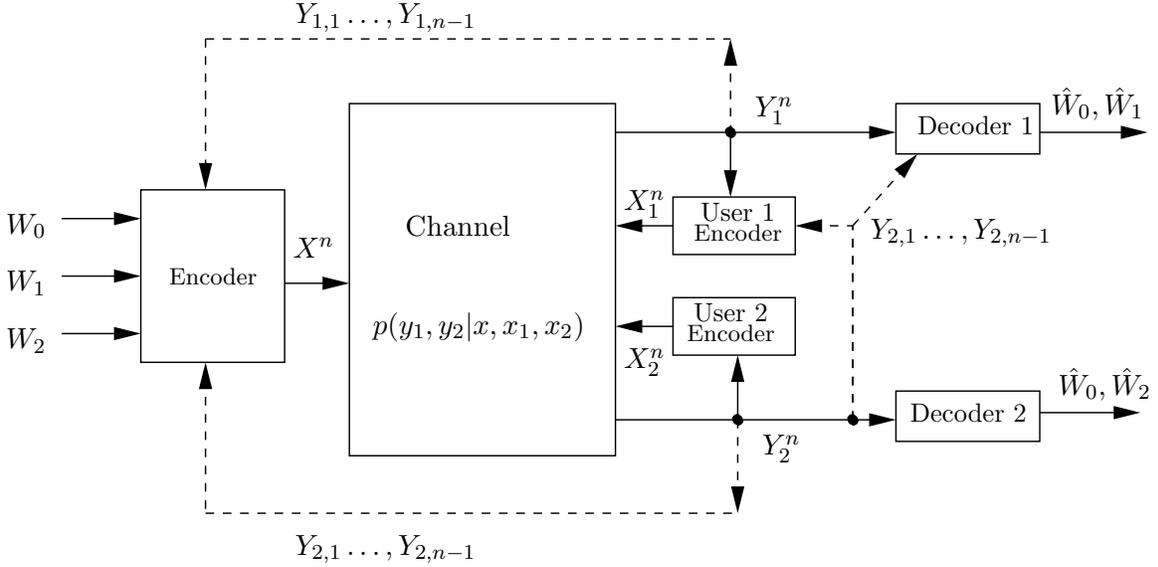

Figure 9: Fully Cooperative RBC with feedback

## 5.1 Discrete Memoryless Fully Cooperative RBCs with Feedback

Since the Fully Cooperative RBC with feedback is degraded, we can obtain the capacity region of this feedback channel.

**Theorem 16.** *(Capacity Region for Fully Cooperative RBC with Feedback) The capacity region of the Fully Cooperative RBC with feedback is given by the rate region with the rate tuples $(R_0, R_1, R_2)$ that satisfy*

$$\begin{aligned} R_0 + R_2 &< \min\left\{I(U, X_1; Y_2 | X_2), I(U; Y_1, Y_2 | X_1, X_2)\right\}, \\ R_1 &< I(X; Y_1, Y_2 \mid U, X_1, X_2) \end{aligned} \tag{44}$$

*for some joint distribution $p(x_1, x_2, u)p(x|u)p(y_1, y_2|x, x_1, x_2)$. The auxiliary random variable $U$ is bounded in cardinality by $|\mathcal{U}| \leq |\mathcal{X}| \cdot |\mathcal{X}_1| \cdot |\mathcal{X}_2| + 2$.*

*Proof.* Theorem 9 provides the achievability with $(Y_1, Y_2)$ replacing $Y_1$ as the output at user 1. The proof of the converse is similar to those for Theorems 2 and 6, and is hence omitted. □

**Remark 7.** *From the proof of the achievability, the source does not exploit feedback information to achieve the capacity region of the channel. Hence even if the channel only has feedback from the user 2 to user 1, the capacity region is still the same as the channel with additional feedback from both users to the source. This is similar to what we have remarked for the Partially cooperative RBC with feedback.*

**Remark 8.** *For the Fully Cooperative RBC, since the output at user 2 is fed back to user 1, it may seem that the relay link from user 2 to user 1 would not serve a useful purpose. This is indeed true for Gaussian channels as we will study in the next subsection. However, this may not be the case for the discrete memoryless channel. The reason is that although user 2 does not need to forward any information to user 1, the relay input sent by user 2 may still affect the channel. Hence user 2 can send a single codeword through the relay link to result in the best achievable region.*

Note that if the original Fully Cooperative RBC is degraded, the capacity region given in Theorem 16 is the same as the capacity region for the same channel without feedback given in Theorem 12.

**Corollary 5.** *Feedback does not increase the capacity region of the degraded Fully Cooperative RBC.*

## 5.2 Gaussian Fully Cooperative RBCs with Feedback

In this section, we study two Gaussian Fully Cooperative RBCs with feedback: the D-AWGN and the AWGN cases.

We first have the following capacity theorem for the D-AWGN Fully Cooperative RBC with feedback.

**Theorem 17.** *(Capacity Region for D-AWGN Fully Cooperative RBC) Feedback does not increase the capacity region for the D-AWGN Fully Cooperative RBC, i.e., the capacity region for the D-AWGN Fully Cooperative RBC with feedback is the same as the capacity region for the D-AWGN Fully/Partially Cooperative RBC, and is given by Theorem 4.*

*Proof.* The proof of the achievability is given by that for the D-AWGN Fully Cooperative RBC without feedback. The proof of the converse is similar to that for the D-AWGN Partially Cooperative RBC with feedback, which is given in Appendix E. □

Note that for the D-AWGN Partially Cooperative RBC, we have studied a Self-Interfered channel, which is a reasonable model from a practical point of view. Similarly, it is of interest to study the following *Self-Interfered D-AWGN Fully Cooperative RBC* model, where the outputs at users 1 and 2 are given by

$$\begin{aligned} Y_1 &= X + aX_1 + bX_2 + Z_1 \\ Y_2 &= X + X_1 + dX_2 + Z_1 + Z' \end{aligned} \quad (45)$$

where $Z_1$ and $Z'$ are independent zero mean real Gaussian random variables with variances $N_1$ and $N_2 - N_1$, respectively, and parameters $a, b$ and $d$ are real numbers.

Note that the Self-Interfered D-AWGN Fully Cooperative RBC is also degraded. Theorem 17 still holds for this channel with feedback.

**Corollary 6.** *The capacity region for the Self-Interfered D-AWGN Fully Cooperative RBC with/without feedback is the same as the capacity region for the D-AWGN Fully Cooperative RBC with/without feedback, and is given in Theorem 17.*

We now consider the AWGN Fully Cooperative RBC with feedback. The following theorem provides the capacity region for this channel.

**Theorem 18.** *(Capacity Region for AWGN Fully Cooperative RBC with Feedback) The capacity region for the AWGN Fully Cooperative RBC with feedback is the same as the capacity region for the AWGN Partially Cooperative RBC with feedback given in Theorem 8, i.e., the capacity region* is given by the rate region with the rate tuples $(R_0, R_1, R_2)$ satisfying

$$R_0 + R_2 < \max_{0 \leq \beta \leq 1} \min \left\{ \mathcal{C}\left(\frac{P_1 + \bar{\alpha}P + 2\sqrt{\bar{\beta}\bar{\alpha}PP_1}}{\alpha P + N_2}\right), \mathcal{C}\left(\frac{\beta\bar{\alpha}P}{\alpha P + \frac{N_1 N_2}{N_1 + N_2}}\right) \right\},$$

$$R_1 < \mathcal{C}\left(\frac{\alpha P}{\frac{N_1 N_2}{N_1 + N_2}}\right)$$

(46)

for some $\alpha \in [0, 1]$.

*Proof.* The proof follows similarly as that for Theorem 8, and is briefly summarized as follows.

We define

$$S := \frac{N_1 Y_2 + N_2 Y_1}{N_1 + N_2}, \tag{47}$$

and the mapping from $(Y_1, Y_2)$ to $(S, Y_2)$ is one-to-one. Hence the channel with the outputs being $(Y_1, Y_2)$ and $Y_2$ is equivalent to the channel with the outputs being $(S, Y_2)$ and $Y_2$.

We express $S$ and $Y_2$ in the following form

$$S = X + \frac{N_1}{N_1 + N_2} X_1 + \frac{N_2}{N_1 + N_2} X_2 + \hat{Z}_1, \tag{48}$$

$$Y_2 = S + \frac{N_2}{N_1 + N_2} X_1 - \frac{N_2}{N_1 + N_2} X_2 + \hat{Z}, \tag{49}$$

where $\hat{Z}_1 := \frac{N_1 Z_2 + N_2 Z_1}{N_1 + N_2}$ and $\hat{Z} := \frac{N_2}{N_1 + N_2}(Z_2 - Z_1)$.

It is clear that $\hat{Z}$ is independent of $X$, $X_1$ and $X_2$, and is also independent of $\hat{Z}_1$. Hence $\hat{Z}$ is independent of $S$. Therefore, $Y_2$ is independent of $X$, given $(S, X_1, X_2)$, i.e., $Y_2$ is a degraded version of $S$.

Now, for the equivalent channel with outputs being $(S, Y_2)$ and $Y_2$, we have

$$S = X + \frac{N_1}{N_1 + N_2} X_1 + \frac{N_2}{N_1 + N_2} X_2 + \hat{Z}_1$$

$$Y_2 = X + X_1 + \hat{Z}_1 + \hat{Z}$$

(50)

where $\hat{Z}_1$ and $\hat{Z}$ are independent zero mean real Gaussian random variables with variances $\frac{N_1 N_2}{N_1 + N_2}$ and $\frac{N_2^2}{N_2 + N_1}$, respectively.

This is a Self-Interfered D-AWGN Fully Cooperative RBC defined in (45) with feedback, and hence Corollary 6 can be applied to yield the capacity region. □

We have the following remarks for the capacity region for the AWGN Fully Cooperative RBC with feedback.

**Remark 9.** *Theorem 18 implies that feedback effectively changes the AWGN Fully Cooperative RBC with feedback to a D-AWGN Fully Cooperative RBC with feedback but with noise variance $N_1$ being replaced by $\frac{N_1 N_2}{N_1 + N_2}$.*

**Remark 10.** *For both D-AWGN and AWGN Fully Cooperative RBCs with feedback, their capacity regions are the same as the corresponding Partially Cooperative RBCs with feedback. Hence, for these two Gaussian channels with feedback, the relay link from user 2 to user 1 does not help, because all the useful information at user 2 has been conveyed to user 1 through feedback.*

We now compare the capacity region of the AWGN Fully Cooperative RBC with feedback given in Theorem 18 with the outer bound on the capacity region of the AWGN Fully Cooperative RBC given in Theorem 15. It is clear that the capacity region of the AWGN Fully Cooperative RBC with feedback contains the outer bound on the capacity region of the AWGN Fully Cooperative RBC without feedback. In particular, for small values of the relay power $P_2$, the capacity region of the AWGN Fully Cooperative RBC with feedback strictly contains the outer bound on the capacity of the same channel without feedback, i.e., feedback enlarges the capacity region for the AWGN Fully Cooperative RBC for these cases.

## 6 Comments on Power Constraints

In Section 2.3, we showed that the achievable region of the AWGN Partially Cooperative RBC (given in Corollary 1) is larger than the capacity region of the original Gaussian broadcast channel. This is also demonstrated by numerical example in Figure 5. However, this comparison is based on the assumption that the power constraint at the source for the AWGN Partially Cooperative RBC is the same as the power constraint at the source for the Gaussian broadcast channel, and that there is an additional power $P_1$ for the relay node (user 1) to transmit relaying information. Hence it is conceivable that the improvement in the capacity region for the AWGN Partially Cooperative RBC is due to this additional power at the relay node. We now consider a case for the AWGN Partially Cooperative RBC where the total power available for the source and relay is the same as the power available for the source in the broadcast channel, i.e., the source and the relay node need to share the amount of power $P$. We explore whether the AWGN Partially Cooperative RBC still has a larger capacity region than the Gaussian broadcast channel.

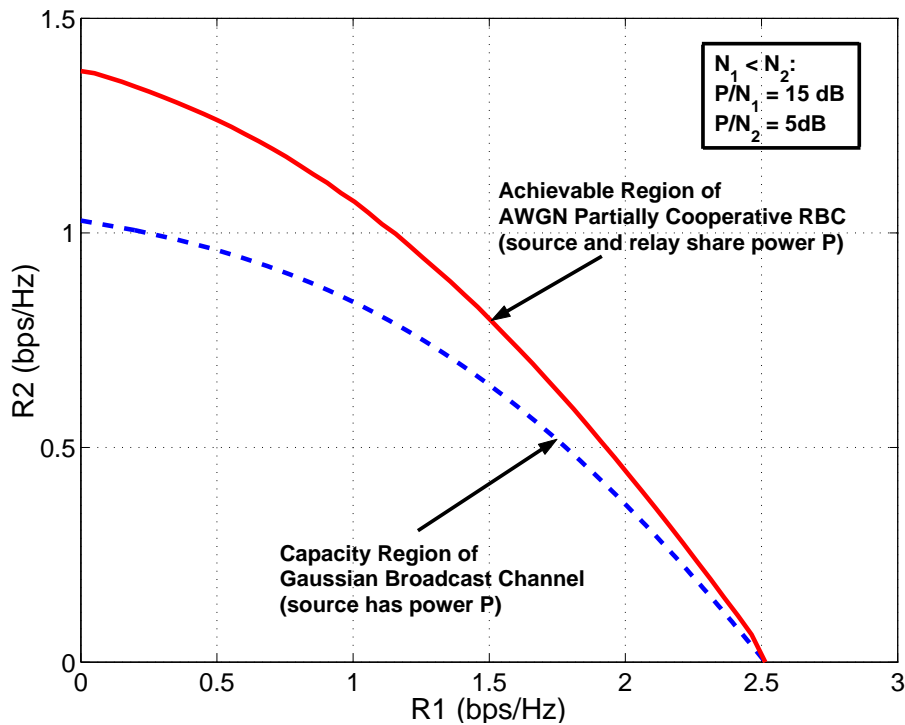

Figure 10: Comparison of rate regions

In Figure 10, we plot the achievable rate region (boundary with solid line) for the AWGN Partially Cooperative RBC, where $N_1 < N_2$, and where the source and the relay node share the amount of power $P$. We compare this achievable rate region with the capacity region (boundary with dashed line) of the Gaussian broadcast channel with the power constraint $P$ for the source. It is clear from the graph that the achievable region of the AWGN Partially Cooperative RBC is larger than the capacity region of the Gaussian broadcast channel. Clearly this enlargement is not due to the additional power at the relay node any more. The gain comes from the fact that the source and relay can coherently transmit information to user 2. Of course this gain due to coherent combining is limited by how much information the source can forward to the relay node before the two nodes can actually cooperate. Hence the stronger the link from the source to relay is, the larger the coherent combining gain that can be achieved. Since we have assumed that $N_1 < N_2$, which means that the source can forward more information to the relay than to user 2 through the direct link from the source to user 2, the coherent combining gain always exists even if it may be small.

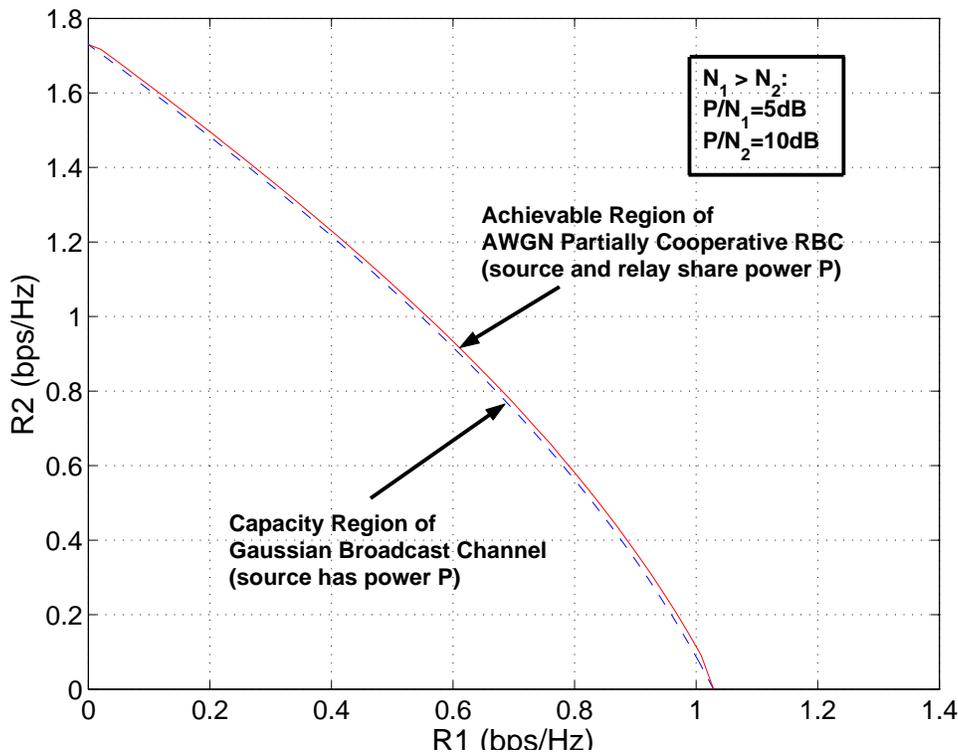

Figure 11: Comparison of rate regions

We next consider the AWGN Partially Cooperative RBC where $N_1 > N_2$, i.e., where the relay has a worse channel from the source than user 2. We note that when the relay has an additional power $P_1$, the achievable region given in (20) is larger than the capacity region of the original Gaussian broadcast channel. Now even when the source and relay are subject to a sum power constraint $P$, the achievable rate region based on the estimate-and-forward scheme for the AWGN Partially Cooperative RBC is larger than the capacity region of the original broadcast channel (see Figure 11). However, this improvement is not due to coherent combining between the source and the relay as in the case where $N_1 < N_2$. The reason for the improvement is that in this case the source needs to split some power for the relay and hence the source signals cause less interference at the relay. Thus the relay is able to decode at at a higher rate. Figure 11 also suggests that this improvement is small, and that the maximum rate of user 2 is not improved. This suggests that in the case where the relay has a worse channel from the source than user 2, the relay transmission may not help much in enlarging the capacity region unless an additional amount of power is available at the relay. On the other hand, in this case letting user 2 be the relay node makes the relaying more helpful, because now the relay (user 2) has a better channel from the source than user 1, and coherent combining helps to enlarge the capacity region.

Similarly, the achievable rate region of the AWGN Fully Cooperative RBC, where the source and two users share the amount of power $P$, is larger than the capacity region of the original Gaussian broadcast channel with the source subject to power constraint $P$. In the AWGN Fully Cooperative RBC, the relay node with a better channel from the source helps more towards enlarging the capacity region due to coherent combining. The relay node with a worse channel from the source helps only a little due to less interference in decoding as in the preceding discussion.

# 7 Concluding Remarks

We performed a comprehensive information-theoretic study of two relay broadcast channels, the Partially Cooperative RBC and the Fully Cooperative RBC. We derived bounds on the capacity region for these channels, and established the capacity region for the special cases of degraded channels. We demonstrated via Gaussian examples that these RBCs have significant gains in capacity region compared to standard broadcast channels. Our results suggest that cooperative relaying is a powerful technique in achieving high speed communication for wireless downlink systems and other networks that include broadcast transmissions.

In analyzing the capacity regions of RBCs, We provided an alternative to the cut-set bound approach to obtain outer bounds. We showed that our outer bounds are tighter than those based on the cut-set bound, and that they are close to the corresponding inner bounds in Gaussian examples. We believe that our technique for deriving these outer bounds is applicable more generally, and may be useful in deriving tighter outer bounds than the cut-set bound in other network information theory problems.

For the RBCs studied in the paper, the relay is allowed to transmit and receive at the same time in the same frequency band. In practice, models where the relays transmit and receive in orthogonal channels may be of interest. These RBCs have been studied in recent papers from an information-theoretic viewpoint [8, 9], and in the fading channel context [31].

In this paper, we have focused purely on the information-theoretic aspect of the RBCs. Further studies on this topic from coding and networking viewpoints will allow for the implementation of relaying and user cooperation in future wireless networks.

# Appendix

# A  Outline of Proof for Theorem 1

We assume that the source uses the superposition coding which is optimal for the degraded broadcast channel [6, Chapter 14.6]. We also assume that the relay (user 1) uses the decode-and-forward relaying scheme [5, Section II]. We adopt the regular encoding/sliding window decoding strategy [2] for the decode-and-forward scheme which is different from the irregular encoding/successive decoding strategy used in [5, Section II]. A review of three decode-and-forward strategies can be found in [24].

We first prove that without common message $W_0$, the following rate pair is achievable:

$$\begin{aligned} R_2 &< \min\left\{I(U, X_1; Y_2), I(U; Y_1 \mid X_1)\right\}, \\ R_1 &< I(X; Y_1 \mid U, X_1). \end{aligned} \qquad (51)$$

Then, from the following proof, it is easily seen that user 1 decodes the messages for both users 1 and 2. We can hence view part of the rate $R_2$ to be the common rate $R_0$, and the rate region given in Theorem 1 is achievable.

We consider a transmission over $B$ blocks, each with length $n$. At each of the first $B-1$ blocks, a message pair $(W_{1,i}, W_{2,i}) \in [1, 2^{nR_1}] \times [1, 2^{nR_2}]$ is encoded and sent from the source, where $i$ denotes the index of the block, and $i = 1, 2, \ldots, B-1$. For fixed $n$, the rate pair $\left(R_1 \frac{B-1}{B}, R_2 \frac{B-1}{B}\right)$ approaches $(R_1, R_2)$ as $B \to \infty$.

We use random codes for the proof. Fix a joint probability distribution of $X_1, U, X, Y_1, Y_2$:

$$p(x_1)p(u|x_1)p(x|u, x_1)p(y_1, y_2|x, x_1)$$

where $U$ is an auxiliary random variable that stands for the information being carried by the source input that is intended for user 2. In the following, we use $A_\epsilon^{(n)}$ to denote the jointly $\epsilon$-typical set (see [6, Chapter 14.2] for definition) based on this joint distribution.

*Random Codebook Generation:* We generate two statistically independent random codebooks 1 and 2 by the following same steps.

1. Generate $2^{nR_2}$ i.i.d. $\underline{x}_1$ each with distribution $\prod_{i=1}^n p(x_{1,i})$. Index $\underline{x}_1(w_2')$, $w_2' \in [1, 2^{nR_2}]$.

2. For each $\underline{x}_1(w_2')$, generate $2^{nR_2}$ i.i.d. $\underline{u}$ each with distribution $\prod_{i=1}^n p(u_i|x_{1,i}(w_2'))$. Index $\underline{u}(w_2', w_2)$, $w_2 \in [1, 2^{nR_2}]$.

3. For each $\underline{x}_1(w_2')$ and $\underline{u}(w_2', w_2)$, generate $2^{nR_1}$ i.i.d. $\underline{x}$ each with distribution $\prod_{i=1}^n p(x_i|u_i(w_2', w_2), x_{1,i}(w_2'))$. Index $\underline{x}(w_2', w_2, w_1)$, $w_1 \in [1, 2^{nR_1}]$.

*Encoding:* We encode messages using codebooks 1 and 2, respectively, for blocks with odd and even indices. This is because some of the following decoding steps are performed jointly over two adjacent blocks, and having independent codebooks makes the error events corresponding to these blocks independent, thus making the probabilities of these error events easy to calculate.

At the beginning of block $i$, let $(w_{1,i}, w_{2,i})$ be the new message pair to be sent from the source in block $i$, and $(w_{1,i-1}, w_{2,i-1})$ be the message pair being sent from the source in previous block $i-1$. The source encoder then sends $\underline{x}(w_{2,i-1}, w_{2,i}, w_{1,i})$.

At the beginning of block $i$, user 1 (relay node) has decoded the message $w_{2,i-1}$ transmitted from the source in previous block $i-1$. It then sends the codeword $\underline{x}_1(w_{2,i-1})$.

For convenience, we list the codewords that are sent in the first three blocks in the following table.

| block 1 | block 2 | block 3 |
|---|---|---|
| $\underline{x}_1(1)$ | $\underline{x}_1(w_{2,1})$ | $\underline{x}_1(w_{2,2})$ |
| $\underline{u}\ (1, w_{2,1})$ | $\underline{u}\ (w_{2,1}, w_{2,2})$ | $\underline{u}\ (w_{2,2}, w_{2,3})$ |
| $\underline{x}\ (1, w_{2,1}, w_{1,1})$ | $\underline{x}\ (w_{2,1}, w_{2,2}, w_{1,2})$ | $\underline{x}\ (w_{2,2}, w_{2,3}, w_{1,3})$ |

*Decoding:* The decoding procedures at the end of block $i$ are as follows.

1. User 1, having known $w_{2,i-1}$, declares the message $\hat{w}_{2,i}$ is sent if there is a unique $\hat{w}_{2,i}$ such that $\left(\underline{x}_1(w_{2,i-1}), \underline{u}(w_{2,i-1}, \hat{w}_{2,i}), \underline{y}_1(i)\right) \in A_\epsilon^{(n)}$. It can be shown that the decoding error in this step is small for sufficiently large $n$ if

$$R_2 < I(U; Y_1|X_1). \tag{52}$$

2. User 1, having known $w_{2,i-1}$ and $w_{2,i}$, declares the message $\hat{w}_{1,i}$ is sent if there is a unique $\hat{w}_{1,i}$ such that $\left(\underline{x}_1(w_{2,i-1}), \underline{u}(w_{2,i-1}, w_{2,i}), \underline{x}(w_{2,i-1}, w_{2,i}, \hat{w}_{1,i}), \underline{y}_1(i)\right) \in A_\epsilon^{(n)}$. It can be shown that the decoding error in this step is small for sufficiently large $n$ if

$$R_1 < I(X; Y_1 \mid U, X_1). \tag{53}$$

3. User 2, having known $w_{2,i-2}$, decodes $w_{2,i-1}$ based on the information received in blocks $i-1$ and $i$. It declares that the message $\hat{\hat{w}}_{2,i-1}$ is sent if there is a unique $\hat{\hat{w}}_{2,i-1}$ such that $(\underline{x}_1(w_{2,i-2}), \underline{u}(w_{2,i-2}, \hat{\hat{w}}_{2,i-1}), \underline{y}_2(i-1)) \in A_\epsilon^{(n)}$ and $(\underline{x}_1(\hat{\hat{w}}_{2,i-1}), \underline{y}_2(i)) \in A_\epsilon^{(n)}$. It can be shown that the decoding error in this step is small for sufficiently large $n$ if

$$R_2 < I(U; Y_2|X_1) + I(X_1; Y_2) = I(X_1, U; Y_2). \tag{54}$$

Combining equations (52), (53) and (54), we conclude that the rate region given in (51) is achievable.

Finally, the cardinality of the auxiliary random variable $U$ can be bounded by applying standard techniques (e.g., see [7, Lemma 3.4]).

# B  Proof of Theorem 2

The proof uses techniques that are used in proving the converse of the capacity region of the degraded broadcast channel [6, Chapter 14, Problem 11], and in proving the upper bound on the capacity region for the relay channel [5, Section III].

We consider a sequence of $\left(2^{nR_0}, 2^{nR_1}, 2^{nR_2}, n\right)$ codes for a Partially Cooperative RBC with $P_e^{(n)} \to 0$. Then the probability distribution on the joint ensemble space $W_0 \times W_1 \times W_2 \times \mathcal{X}^n \times \mathcal{X}_1^n \times \mathcal{Y}_1^n \times \mathcal{Y}_2^n$ is given by

$$\begin{aligned}&p(w_0, w_1, w_2, x^n, x_1^n, y_1^n, y_2^n) \\ &= p(w_0)p(w_1)p(w_2)p(x^n|w_0, w_1, w_2)\prod_{i=1}^n p(x_{1,i}|y_1^{i-1})p(y_{1,i}, y_{2,i}|x_i, x_{1,i}).\end{aligned} \tag{55}$$

By Fano's Inequality, we have

$$\begin{aligned}H(W_1|Y_1^n) &\leq H(W_0, W_1|Y_1^n) \leq n(R_0 + R_1)P_e^{(n)} + 1 := n\delta_{1,n} \\ H(W_2|Y_2^n) &\leq H(W_0, W_2|Y_2^n) \leq n(R_0 + R_2)P_e^{(n)} + 1 := n\delta_{2,n}\end{aligned} \tag{56}$$

Note that $\delta_{1,n}, \delta_{2,n} \to 0$ if $P_e^{(n)} \to 0$.

We first consider

$$\begin{aligned}
nR_0 + nR_2 &= H(W_0, W_2) = I(W_0, W_2; Y_2^n) + H(W_0, W_2 | Y_2^n) \\
&\leq I(W_0, W_2; Y_2^n) + n\delta_{2,n} \\
&= \sum_{i=1}^n I(W_0, W_2; Y_{2,i} | Y_2^{i-1}) + n\delta_{2,n} \\
&= \sum_{i=1}^n H(Y_{2,i} | Y_2^{i-1}) - H(Y_{2,i} | Y_2^{i-1}, W_0, W_2) + n\delta_{2,n} \\
&\leq \sum_{i=1}^n H(Y_{2,i}) - H(Y_{2,i} | Y_2^{i-1}, W_0, W_2, Y_1^{i-1}, X_{1,i}) + n\delta_{2,n} \\
&= \sum_{i=1}^n H(Y_{2,i}) - H(Y_{2,i} | U_i, X_{1,i}) + n\delta_{2,n} \\
&= \sum_{i=1}^n I(U_i, X_{1,i}; Y_{2,i}) + n\delta_{2,n}
\end{aligned} \quad (57)$$

where we defined $U_i := (W_0, W_2, Y_1^{i-1}, Y_2^{i-1})$. Note that $X_{1,i} \to U_i \to X_i$ form a Markov chain, and $U_i \to (X_i, X_{1,i}) \to (Y_{1,i}, Y_{2,i})$ also form a Markov chain.

We then consider

$$\begin{aligned}
nR_0 + nR_2 &\leq I(W_0, W_2; Y_2^n) + n\delta_{2,n} \leq I(W_0, W_2; Y_2^n, Y_1^n) + n\delta_{2,n} \\
&\stackrel{(p)}{=} \sum_{i=1}^n I(W_0, W_2; Y_{2,i}, Y_{1,i} | Y_2^{i-1}, Y_1^{i-1}) + n\delta_{2,n} \\
&= \sum_{i=1}^n H(W_0, W_2 | Y_2^{i-1}, Y_1^{i-1}) - H(W_0, W_2 | Y_2^i, Y_1^i) + n\delta_{2,n} \\
&\stackrel{(a)}{\leq} \sum_{i=1}^n H(W_0, W_2 | Y_2^{i-1}, Y_1^{i-1}, X_{1,i}) - H(W_0, W_2 | Y_2^i, Y_1^i, X_{1,i}) + n\delta_{2,n} \\
&\stackrel{(q)}{=} \sum_{i=1}^n I(W_0, W_2; Y_{1,i}, Y_{2,i} | Y_1^{i-1}, Y_2^{i-1}, X_{1,i}) + n\delta_{2,n} \\
&= \sum_{i=1}^n H(Y_{1,i}, Y_{2,i} | Y_1^{i-1}, Y_2^{i-1}, X_{1,i}) - H(Y_{1,i}, Y_{2,i} | Y_1^{i-1}, Y_2^{i-1}, X_{1,i}, W_0, W_2) + n\delta_{2,n} \\
&\leq \sum_{i=1}^n H(Y_{1,i}, Y_{2,i} | X_{1,i}) - H(Y_{1,i}, Y_{2,i} | U_i, X_{1,i}) + n\delta_{2,n} \\
&= \sum_{i=1}^n I(U_i; Y_{1,i}, Y_{2,i} | X_{1,i}) + n\delta_{2,n}
\end{aligned} \quad (58)$$

where $(a)$ follows from the fact that conditioned on $(Y_2^{i-1}, Y_1^{i-1})$, $X_{1,i}$ is independent of $W_0, W_2$.

We next consider

$$
\begin{aligned}
nR_1 &= H(W_1) = I(W_1; Y_1^n) + H(W_1|Y_1^n) \leq I(W_1; Y_1^n, Y_2^n, W_0, W_2) + n\delta_{1,n} \\
&= I(W_1; Y_1^n, Y_2^n | W_0, W_2) + n\delta_{1,n} \\
&= \sum_{i=1}^n I(W_1; Y_{1,i}, Y_{2,i} | Y_1^{i-1}, Y_2^{i-1}, W_0, W_2) + n\delta_{1,n} \\
&\stackrel{(b)}{\leq} \sum_{i=1}^n I(W_1; Y_{1,i}, Y_{2,i} | Y_1^{i-1}, Y_2^{i-1}, W_0, W_2, X_{1,i}) + n\delta_{1,n} \\
&\leq \sum_{i=1}^n H(Y_{1,i}, Y_{2,i} | U_i, X_{1,i}) - H(Y_{1,i}, Y_{2,i} | Y_1^{i-1}, Y_2^{i-1}, W_0, W_2, W_1, X_{1,i}, X_i) + n\delta_{1,n} \quad (59) \\
&= \sum_{i=1}^n H(Y_{1,i}, Y_{2,i} | U_i, X_{1,i}) - H(Y_{1,i}, Y_{2,i} | Y_1^{i-1}, Y_2^{i-1}, W_0, W_2, X_{1,i}, X_i) + n\delta_{1,n} \\
&= \sum_{i=1}^n H(Y_{1,i}, Y_{2,i} | U_i, X_{1,i}) - H(Y_{1,i}, Y_{2,i} | U_i, X_{1,i}, X_i) + n\delta_{1,n} \\
&= \sum_{i=1}^n I(X_i; Y_{1,i}, Y_{2,i} | U_i, X_{1,i}) + n\delta_{1,n}
\end{aligned}
$$

where $(b)$ follows from the same reasoning as in the steps from $(p)$ to $(q)$ in (58).

We now consider

$$
\begin{aligned}
nR_0 + nR_1 &= H(W_0, W_1) = I(W_0, W_1; Y_1^n) + H(W_0, W_1 | Y_1^n) \\
&\leq I(W_0, W_1; Y_1^n) + n\delta_{1,n} \\
&= \sum_{i=1}^n I(W_0, W_1; Y_{1,i} | Y_1^{i-1}) + n\delta_{1,n} \\
&\stackrel{(c)}{\leq} \sum_{i=1}^n I(W_0, W_1; Y_{1,i} | Y_1^{i-1}, X_{1,i}) + n\delta_{1,n} \\
&\leq \sum_{i=1}^n H(Y_{1,i} | Y_1^{i-1}, X_{1,i}) - H(Y_{1,i} | W_0, W_1, Y_1^{i-1}, Y_2^{i-1}, X_{1,i}) + n\delta_{1,n} \\
&\leq \sum_{i=1}^n H(Y_{1,i} | X_{1,i}) - H(Y_{1,i} | U_i', X_{1,i}) + n\delta_{1,n} \\
&\leq \sum_{i=1}^n I(U_i'; Y_{1,i} | X_{1,i}) + n\delta_{1,n}
\end{aligned} \quad (60)
$$

where $(c)$ follows from the same reasoning as in the steps from $(p)$ to $(q)$ in (58). We defined $U_i' := (W_0, W_1, Y_1^{i-1}, Y_2^{i-1})$. Note that $X_{1,i} \to U_i' \to X_i$ form a Markov chain, and $U_i' \to (X_i, X_{1,i}) \to (Y_{1,i}, Y_{2,i})$ also form a Markov chain.

We finally consider

$$\begin{aligned}
nR_2 &= H(W_2) = I(W_2;Y_2^n) + H(W_2|Y_2^n) \\
&\leq I(W_2;Y_2^n) + n\delta_{2,n} \leq I(W_2;Y_1^n,Y_2^n,W_0,W_1) + n\delta_{2,n} \\
&= I(W_2;Y_1^n,Y_2^n|W_0,W_1) + n\delta_{2,n} \\
&= \sum_{i=1}^n I(W_2;Y_{1,i},Y_{2,i}|Y_1^{i-1},Y_2^{i-1},W_0,W_1) + n\delta_{2,n} \\
&\stackrel{(d)}{\leq} \sum_{i=1}^n I(W_2;Y_{1,i},Y_{2,i}|Y_1^{i-1},Y_2^{i-1},W_0,W_1,X_{1,i}) + n\delta_{2,n} \\
&\leq \sum_{i=1}^n H(Y_{1,i},Y_{2,i}|U'_i,X_{1,i}) - H(Y_{1,i},Y_{2,i}|Y_1^{i-1},Y_2^{i-1},W_0,W_1,W_2,X_{1,i},X_i) + n\delta_{2,n} \\
&= \sum_{i=1}^n H(Y_{1,i},Y_{2,i}|U'_i,X_{1,i}) - H(Y_{1,i},Y_{2,i}|U'_i,X_{1,i},X_i) + n\delta_{2,n} \\
&= \sum_{i=1}^n I(X_i;Y_{1,i},Y_{2,i}|U'_i,X_{1,i}) + n\delta_{2,n}
\end{aligned} \qquad (61)$$

where $(d)$ follows from the same reasoning as in the steps from $(p)$ to $(q)$ in (58).

Now in order to change the upper bounds that we have derived in (57)-(61) to single letter characterizations, we introduce a random variable $Q$ which is independent of $W_0, W_1, W_2, X^n, X_1^n, Y_1^n, Y_2^n$, and is uniformly distributed over $\{1,2,\ldots,n\}$. Define $U = (Q, U_Q)$, $U' = (Q, U'_Q)$, $X = X_Q$, $X_1 = X_{1,Q}$, $Y_1 = Y_{1,Q}$, and $Y_2 = Y_{2,Q}$. Clearly, we have Markov chains: $X_1 \to U \to X$, $X_1 \to U' \to X$, and $(U,U') \to (X,X_1) \to (Y_1,Y_2)$. By using the above definitions, equations (57)−(61), become

$$\begin{aligned}
R_0 + R_2 &\leq \frac{1}{n}\sum_{i=1}^n I(U_i,X_{1,i};Y_{2,i}) + \delta_{2,n} = I(U_Q,X_{1,Q};Y_{2,Q}|Q) + \delta_{2,n} \\
&\leq I(Q,U_Q,X_{1,Q};Y_{2,Q}) + \delta_{2,n} = I(U,X_1;Y_2) + \delta_{2,n}
\end{aligned} \qquad (62)$$

$$\begin{aligned}
R_0 + R_2 &\leq \frac{1}{n}\sum_{i=1}^n I(U_i;Y_{1,i},Y_{2,i}|X_{1,i}) + \delta_{2,n} = I(U_Q;Y_{1,Q},Y_{2,Q}|X_{1,Q},Q) + \delta_{2,n} \\
&\leq I(Q,U_Q;Y_{1,Q},Y_{2,Q}|X_{1,Q}) + \delta_{2,n} = I(U;Y_1,Y_2|X_1) + \delta_{2,n}
\end{aligned} \qquad (63)$$

$$\begin{aligned}
R_1 &\leq \frac{1}{n}\sum_{i=1}^n I(X_i;Y_{1,i},Y_{2,i}|U_i,X_{1,i}) + \delta_{1,n} = I(X_Q;Y_{1,Q},Y_{2,Q}|U_Q,X_{1,Q},Q) + \delta_{1,n} \\
&= I(X;Y_1,Y_2|U,X_1) + \delta_{1,n}
\end{aligned} \qquad (64)$$

$$\begin{aligned}
R_0 + R_1 &\leq \frac{1}{n}\sum_{i=1}^n I(U'_i;Y_{1,i}|X_{1,i}) + \delta_{1,n} = I(U'_Q;Y_{1,Q}|X_{1,Q},Q) + \delta_{1,n} \\
&\leq I(Q,U'_Q;Y_{1,Q}|X_{1,Q}) + \delta_{1,n} = I(U';Y_1|X_1) + \delta_{1,n}
\end{aligned} \qquad (65)$$

$$\begin{aligned}
R_2 &\leq \frac{1}{n}\sum_{i=1}^n I(X_i;Y_{1,i},Y_{2,i}|U'_i,X_{1,i}) + \delta_{2,n} = I(X_Q;Y_{1,Q},Y_{2,Q}|U'_Q,X_{1,Q},Q) + \delta_{2,n} \\
&= I(X;Y_1,Y_2|U',X_1) + \delta_{2,n}
\end{aligned} \qquad (66)$$

## C  Proof of the Converse for Theorem 3

We only need to add the following steps for equations (58) and (59) in Appendix B by taking into account of the degradedness condition defined in Definition 3. For equation (58), we have

$$nR_0 + nR_2 \leq \sum_{i=1}^{n} I(U_i; Y_{1,i}, Y_{2,i}|X_{1,i}) + n\delta_{2,n}$$
$$= \sum_{i=1}^{n} I(U_i; Y_{1,i}|X_{1,i}) + I(U_i; Y_{2,i}|X_{1,i}, Y_{1,i}) + n\delta_{2,n} \qquad (67)$$
$$= \sum_{i=1}^{n} I(U_i; Y_{1,i}|X_{1,i}) + n\delta_{2,n}$$

where $I(U_i; Y_{2,i}|X_{1,i}, Y_{1,i}) = 0$ follows from the degradedness condition given in Definition 3.

For equation (59), we have

$$nR_1 \leq \sum_{i=1}^{n} I(X_i; Y_{1,i}, Y_{2,i}|U_i, X_{1,i}) + n\delta_{1,n}$$
$$= \sum_{i=1}^{n} I(X_i; Y_{1,i}|U_i, X_{1,i}) + I(X_i; Y_{2,i}|U_i, X_{1,i}, Y_{1,i}) + n\delta_{1,n} \qquad (68)$$
$$= \sum_{i=1}^{n} I(X_i; Y_{1,i}|U_i, X_{1,i}) + n\delta_{1,n}$$

where $I(X_i; Y_{2,i}|U_i, X_{1,i}, Y_{1,i}) = 0$ also follows from the degradedness condition given in Definition 3.

Therefore, (57) in Appendix B, (67) and (68) constitute the converse for the degraded Partially Cooperative RBC.

## D  Proof of the Converse for Theorem 6

The proof is similar to that for Theorem 2 given in Appendix B. We hence provide only an outline, which will be useful in the following proof for the Gaussian case in Appendix E.

We consider a sequence of $\left(2^{nR_0}, 2^{nR_1}, 2^{nR_2}, n\right)$ code with $P_e^{(n)} \to 0$. Then the probability distribution on the joint ensemble space $W_0 \times W_1 \times W_2 \times \mathcal{X}^n \times \mathcal{X}_1^n \times \mathcal{Y}_1^n \times \mathcal{Y}_2^n$ is given by

$$p(w_0, w_1, w_2, x^n, x_1^n, y_1^n, y_2^n)$$
$$= p(w_0)p(w_1)p(w_2) \prod_{i=1}^{n} p(x_i|w_0, w_1, w_2, y_1^{i-1}, y_2^{i-1}) p(x_{1,i}|y_1^{i-1}, y_2^{i-1}) p(y_{1,i}, y_{2,i}|x_i, x_{1,i}). \qquad (69)$$

By Fano's Inequality, we have

$$H(W_1|Y_1^n, Y_2^n) \leq H(W_0, W_1|Y_1^n, Y_2^n) \leq n(R_0 + R_1)P_e^{(n)} + 1 := n\delta_{1,n}$$
$$H(W_0, W_2|Y_2^n) \leq n(R_0 + R_2)P_e^{(n)} + 1 := n\delta_{2,n} \qquad (70)$$

where $\delta_{1,n}, \delta_{2,n} \to 0$ if $P_e^{(n)} \to 0$.

The proof for the following two bounds follows steps that are identical to those in equations (57) and (58) in Appendix B.

$$nR_0 + nR_2 \leq \sum_{i=1}^{n} I(U_i, X_{1,i}; Y_{2,i}) + n\delta_{2,n}. \tag{71}$$

$$nR_0 + nR_2 \leq \sum_{i=1}^{n} I(U_i; Y_{1,i}, Y_{2,i}|X_{1,i}) + n\delta_{2,n} \tag{72}$$

We now consider

$$\begin{aligned}
nR_1 &= H(W_1) = I(W_1; Y_1^n, Y_2^n) + H(W_1|Y_1^n, Y_2^n) \\
&\leq I(W_1; Y_1^n, Y_2^n, W_0, W_2) + n\delta_{1,n} \\
&\leq \sum_{i=1}^{n} I(X_i; Y_{1,i}, Y_{2,i}|U_i, X_{1,i}) + n\delta_{1,n}
\end{aligned} \tag{73}$$

where we omitted the steps that are identical to those in equation (59) in Appendix B.

## E  Proof of the Converse for Theorem 7

The techniques in the proof of the converse for the capacity of the physically degraded Gaussian relay channel [5, Section IV] are useful here, but are not sufficient. In particular, the parameters $\alpha$ and $\beta$ need to be carefully chosen. Moreover, this proof applies the entropy power inequality to the components of the two random vectors, which is different from applying the entropy power inequality to two independent random vectors as in the proof of the converse for the capacity region of the degraded Gaussian broadcast channel. A similar idea has been used in establishing the capacity region of the physically degraded Gaussian broadcast channel with feedback [11].

For the D-AWGN Partially Cooperative RBC, the power constraints at the source and relay imply that the codewords satisfy

$$\sum_{i=1}^{n} \mathsf{E} X_i^2 \leq nP, \qquad \sum_{i=1}^{n} \mathsf{E} X_{1,i}^2 \leq nP_1. \tag{74}$$

We apply the degradedness condition to the bounds (71), (72) and (73) in Appendix D and obtain the following bounds:

$$nR_0 + nR_2 \leq \sum_{i=1}^{n} I(U_i, X_{1,i}; Y_{2,i}) + n\delta_{2,n}, \tag{75}$$

$$nR_0 + nR_2 \leq \sum_{i=1}^{n} I(U_i; Y_{1,i}|X_{1,i}) + n\delta_{2,n}, \tag{76}$$

$$nR_1 \leq \sum_{i=1}^{n} I(X_i; Y_{1,i}|U_i, X_{1,i}) + n\delta_{1,n}. \tag{77}$$

We now apply the bounds (75)–(77) for the D-AWGN Partially Cooperative RBC. We start with (76), and obtain

$$nR_0 + nR_2 \leq \sum_{i=1}^{n} h(Y_{1,i}|X_{1,i}) - h(Y_{1,i}|X_{1,i}, U_i) + n\delta_{2,n}. \tag{78}$$

For the second term in the sum in (78), we have

$$\sum_{i=1}^{n} h(Y_{1,i}|X_{1,i}, U_i) \leq \sum_{i=1}^{n} h(Y_{1,i}) = \sum_{i=1}^{n} h(X_i + Z_{1,i}) \leq \sum_{i=1}^{n} \frac{1}{2} \log 2\pi e(\mathsf{E}X_i^2 + N_1)$$
$$\leq \frac{n}{2} \log 2\pi e \left( \frac{1}{n} \sum_{i=1}^{n} \mathsf{E}X_i^2 + N_1 \right) \leq \frac{n}{2} \log 2\pi e(P + N_1) \tag{79}$$

On the other hand,

$$\sum_{i=1}^{n} h(Y_{1,i}|X_{1,i}, U_i) \geq \sum_{i=1}^{n} h(Y_{1,i}|X_{1,i}, U_i, X_i) = \sum_{i=1}^{n} h(Y_{1,i}|X_i) = \frac{n}{2} \log 2\pi e N_1 \tag{80}$$

where we used that given $X_i$, $Y_{1,i}$ is independent of $X_{1,i}, U_i$.

Combining (79) and (80), we establish that there exists some $\alpha \in [0, 1]$ such that

$$\sum_{i=1}^{n} h(Y_{1,i}|X_{1,i}, U_i) = \frac{n}{2} \log 2\pi e(\alpha P + N_1) \tag{81}$$

For the first term in the sum in (78)

$$\sum_{i=1}^{n} h(Y_{1,i}|X_{1,i}) = \sum_{i=1}^{n} \mathsf{E}h(Y_{1,i}|X_{1,i}) \leq \sum_{i=1}^{n} \mathsf{E}\frac{1}{2} \log 2\pi e \mathsf{Var}(Y_{1,i}|X_{1,i})$$
$$\leq \frac{1}{2} \sum_{i=1}^{n} \log 2\pi e \mathsf{E} \, \mathsf{Var}(X_i + Z_{1,i}|X_{1,i})$$
$$= \frac{1}{2} \sum_{i=1}^{n} \log 2\pi e(\mathsf{E}\mathsf{Var}(X_i|X_{1,i}) + N_1)$$
$$\leq \frac{n}{2} \log 2\pi e \left( \frac{1}{n} \sum_{i=1}^{n} \mathsf{E}\mathsf{Var}(X_i|X_{1,i}) + N_1 \right)$$
$$= \frac{n}{2} \log 2\pi e \left( \frac{1}{n} \sum_{i=1}^{n} \left[ \mathsf{E}\left(X_i^2\right) - \mathsf{E}\left(\mathsf{E}^2(X_i|X_{1,i})\right) \right] + N_1 \right)$$
$$\leq \frac{n}{2} \log 2\pi e \left( P - \frac{1}{n} \sum_{i=1}^{n} \mathsf{E}\left(\mathsf{E}^2(X_i|X_{1,i})\right) + N_1 \right) \tag{82}$$

On the other hand, we know that

$$\sum_{i=1}^{n} h(Y_{1,i}|X_{1,i}) \geq \sum_{i=1}^{n} h(Y_{1,i}|X_{1,i}, U_i) = \frac{n}{2} \log 2\pi e(\alpha P + N_1) \tag{83}$$

where we have used equation (81).

Combining (82) and (83), we obtain

$$\alpha P + N_1 \leq P - \frac{1}{n}\sum_{i=1}^{n} \mathsf{E}\left(\mathsf{E}^2(X_i|X_{1,i})\right) + N_1$$

$$\Rightarrow \quad \frac{1}{n}\sum_{i=1}^{n} \mathsf{E}\left(\mathsf{E}^2(X_i|X_{1,i})\right) \leq \bar{\alpha}P \tag{84}$$

where $\bar{\alpha} = 1 - \alpha$. Hence, there exists some $\beta \in [0,1]$ such that

$$\frac{1}{n}\sum_{i=1}^{n} \mathsf{E}\left(\mathsf{E}^2(X_i|X_{1,i})\right) = \bar{\beta}\bar{\alpha}P \tag{85}$$

where $\bar{\beta} = 1 - \beta$.

We plug the preceding equation in (82), and obtain

$$\sum_{i=1}^{n} h(Y_{1,i}|X_{1,i}) \leq \frac{n}{2}\log 2\pi e\left(P - \bar{\beta}\bar{\alpha}P + N_1\right) = \frac{n}{2}\log 2\pi e\left(\alpha P + \beta\bar{\alpha}P + N_1\right) \tag{86}$$

We plug (81) and (86) in (78), and obtain

$$nR_0 + nR_2 \leq \frac{n}{2}\log 2\pi e\left(\alpha P + \beta\bar{\alpha}P + N_1\right) - \frac{n}{2}\log 2\pi e\left(\alpha P + N_1\right) + n\delta_{2,n}$$

$$= \frac{n}{2}\log\left(1 + \frac{\beta\bar{\alpha}P}{\alpha P + N_1}\right) + n\delta_{2,n} \tag{87}$$

We next consider the bound (75), and have

$$nR_0 + nR_2 \leq \sum_{i=1}^{n} h(Y_{2,i}) - h(Y_{2,i}|U_i, X_{1,i}) + n\delta_{2,n}. \tag{88}$$

The first term in the sum in (88) can be bounded

$$\sum_{i=1}^{n} h(Y_{2,i}) = \sum_{i=1}^{n} h(X_i + X_{1,i} + Z_{1,i} + Z'_i)$$

$$\leq \sum_{i=1}^{n} \frac{1}{2}\log 2\pi e\left(\mathsf{E}(X_i + X_{1,i})^2 + N_2\right) \tag{89}$$

$$\leq \frac{n}{2}\log 2\pi e\left(\frac{1}{n}\sum_{i=1}^{n} \mathsf{E}(X_i + X_{1,i})^2 + N_2\right)$$

For the sum in the preceding equation, we have

$$\frac{1}{n}\sum_{i=1}^{n} \mathsf{E}(X_i + X_{1,i})^2 = \frac{1}{n}\sum_{i=1}^{n} \mathsf{E}X_i^2 + \frac{1}{n}\sum_{i=1}^{n} \mathsf{E}X_{1,i}^2 + \frac{2}{n}\sum_{i=1}^{n} \mathsf{E}X_i X_{1,i}$$

$$\leq P + P_1 + \frac{2}{n}\sum_{i=1}^{n} \mathsf{E}\left(X_{1,i}\mathsf{E}(X_i|X_{1,i})\right)$$

$$\leq P + P_1 + \frac{2}{n}\sum_{i=1}^{n} \sqrt{\mathsf{E}X_{1,i}^2 \cdot \mathsf{E}\left(\mathsf{E}^2(X_i|X_{1,i})\right)} \quad (90)$$

$$\leq P + P_1 + 2\sqrt{\left(\frac{1}{n}\sum_{i=1}^{n} \mathsf{E}X_{1,i}^2\right) \cdot \left(\frac{1}{n}\sum_{i=1}^{n} \mathsf{E}\left(\mathsf{E}^2(X_i|X_{1,i})\right)\right)}$$

$$\leq P + P_1 + 2\sqrt{P_1 \bar{\beta}\bar{\alpha} P}$$

Hence we obtain

$$\sum_{i=1}^{n} h(Y_{2,i}) \leq \frac{n}{2}\log 2\pi e\left(P + P_1 + 2\sqrt{P_1 \bar{\beta}\bar{\alpha} P} + N_2\right) \quad (91)$$

The second term in the sum in (88) can be expressed as

$$\sum_{i=1}^{n} h(Y_{2,i}|U_i, X_{1,i}) = \sum_{i=1}^{n} h(Y_{1,i} + X_{1,i} + Z_i'|U_i, X_{1,i})$$

$$= \sum_{i=1}^{n} h(Y_{1,i} + Z_i'|U_i, X_{1,i}) \quad (92)$$

$$= \sum_{i=1}^{n} \mathsf{E}h(Y_{1,i} + Z_i'|U_i = u_i, X_{1,i} = x_{1,i})$$

We now use the entropy power inequality, and obtain

$$2^{2h(Y_{1,i}+Z_i'|U_i=u_i, X_{1,i}=x_{1,i})} \geq 2^{2h(Y_{1,i}|U_i=u_i, X_{1,i}=x_{1,i})} + 2^{2h(Z_i'|U_i=u_i, X_{1,i}=x_{1,i})}$$

$$= 2^{2h(Y_{1,i}|U_i=u_i, X_{1,i}=x_{1,i})} + 2\pi e(N_2 - N_1)$$

Then,

$$h(Y_{1,i} + Z_i'|U_i = u_i, X_{1,i} = x_{1,i}) \geq \frac{1}{2}\log\left(2^{2h(Y_{1,i}|U_i=u_i, X_{1,i}=x_{1,i})} + 2\pi e(N_2 - N_1)\right)$$

Hence,

$$\mathsf{E}h(Y_{1,i} + Z_i'|U_i = u_i, X_{1,i} = x_{1,i}) \geq \frac{1}{2}\mathsf{E}\log\left(2^{2h(Y_{1,i}|U_i=u_i, X_{1,i}=x_{1,i})} + 2\pi e(N_2 - N_1)\right)$$

$$\geq \frac{1}{2}\log\left(2^{2\mathsf{E}h(Y_{1,i}|U_i=u_i, X_{1,i}=x_{1,i})} + 2\pi e(N_2 - N_1)\right)$$

$$\geq \frac{1}{2}\log\left(2^{2h(Y_{1,i}|U_i, X_{1,i})} + 2\pi e(N_2 - N_1)\right)$$

where we used the fact that $\log(2^x + c)$ is a convex function.

We plug the preceding equation in (92), and obtain

$$\sum_{i=1}^{n} h(Y_{2,i}|U_i, X_{1,i}) \geq \frac{1}{2}\sum_{i=1}^{n} \log\left(2^{2h(Y_{1,i}|U_i,X_{1,i})} + 2\pi e(N_2 - N_1)\right)$$

$$\stackrel{(a)}{\geq} \frac{n}{2}\log\left(2^{2\frac{1}{n}\sum_{i=1}^{n}h(Y_{1,i}|U_i,X_{1,i})} + 2\pi e(N_2 - N_1)\right) \quad (93)$$

$$\stackrel{(b)}{=} \frac{n}{2}\log\left(2\pi e(\alpha P + N_1) + 2\pi e(N_2 - N_1)\right)$$

$$= \frac{n}{2}\log\left(2\pi e(\alpha P + N_2)\right)$$

where $(a)$ also follows from the fact that $\log(2^x + c)$ is a convex function, and $(b)$ follows from (81).

We plug (91) and (93) in (88), and obtain

$$nR_0 + nR_2 \leq \frac{n}{2}\log\frac{P + P_1 + 2\sqrt{\bar{\beta}\bar{\alpha}PP_1} + N_2}{\alpha P + N_2} + n\delta_{2,n}$$

$$= \frac{n}{2}\log\left(1 + \frac{\bar{\alpha}P + P_1 + 2\sqrt{\bar{\beta}\bar{\alpha}PP_1}}{\alpha P + N_2}\right) + n\delta_{2,n} \quad (94)$$

We now consider (77), and have

$$nR_1 \leq \sum_{i=1}^{n} h(Y_{1,i}|U_i, X_{1,i}) - h(Y_{1,i}|U_i, X_{1,i}, X_i) + n\delta_{1,n}$$

$$= \sum_{i=1}^{n} h(Y_{1,i}|U_i, X_{1,i}) - h(Y_{1,i}|X_i) + n\delta_{1,n} \quad (95)$$

$$= \frac{n}{2}\log 2\pi e(\alpha P + N_1) - \frac{n}{2}\log(2\pi e N_1) + n\delta_{1,n}$$

$$= \frac{n}{2}\log\left(1 + \frac{\alpha P}{N_1}\right) + n\delta_{1,n}$$

Therefore, (87), (94) and (95) provide the converse for Theorem 7.

# F  Outline of Proof for Theorem 5

The proof uses the techniques in proving Theorem 8 and techniques in Appendix E.

In the proof for Theorem 8, we have shown that the mapping from $(Y_1, Y_2)$ to $(S, Y_2)$ is one-to-one, where

$$S = X + \frac{N_1}{N_1 + N_2}X_1 + \hat{Z}_1$$
$$Y_2 = X + X_1 + \hat{Z}_1 + \hat{Z} \quad (96)$$

where $\hat{Z}_1$ and $\hat{Z}$ are independent zero mean real Gaussian random variables with variances $\frac{N_1 N_2}{N_1 + N_2}$ and $\frac{N_2^2}{N_2 + N_1}$, respectively. We have also shown that given $(S, X_1)$, $Y_2$ is independent of $X$.

From equations (57)–(60) in Appendix B, we have the following upper bounds:

$$nR_0 + nR_2 = \sum_{i=1}^{n} I(U_i, X_{1,i}; Y_{2,i}) + n\delta_{2,n} \tag{97}$$

$$nR_0 + nR_2 \leq \sum_{i=1}^{n} I(U_i; Y_{1,i}, Y_{2,i}|X_{1,i}) + n\delta_{2,n} \stackrel{(a)}{=} \sum_{i=1}^{n} I(U_i; S_i, Y_{2,i}|X_{1,i}) + n\delta_{2,n}$$
$$\stackrel{(b)}{=} \sum_{i=1}^{n} I(U_i; S_i|X_{1,i}) + n\delta_{2,n} \tag{98}$$

$$nR_1 \leq \sum_{i=1}^{n} I(X_i; Y_{1,i}, Y_{2,i}|U_i, X_{1,i}) + n\delta_{1,n} \stackrel{(c)}{=} \sum_{i=1}^{n} I(X_i; S_i, Y_{2,i}|U_i, X_{1,i}) + n\delta_{1,n}$$
$$\stackrel{(d)}{=} \sum_{i=1}^{n} I(X_i; S_i|U_i, X_{1,i}) + n\delta_{1,n} \tag{99}$$

$$nR_0 + nR_1 \leq \sum_{i=1}^{n} I(U'_i; Y_{1,i}|X_{1,i}) + n\delta_{1,n} \tag{100}$$

where (a) and (c) follow from the fact that the mapping from $(Y_{1,i}, Y_{2,i})$ to $(S_i, Y_{2,i})$ is one-to-one, and (b) and (d) follow from the fact that given $S_i$ and $X_{1,i}$, $Y_{2,i}$ is independent of $U_i$ and $X_i$.

We further bound the equations (97), (98) and (99) by following similar steps in bounding (75), (76) and (77) in Appendix E with $Y_1$ being replaced by $S$, and $N_1$ being replaced by $\frac{N_1 N_2}{N_1+N_2}$. We then obtain the following bounds:

$$nR_0 + nR_2 < \frac{n}{2} \log\left(1 + \frac{P_1 + \bar{\alpha}P + 2\sqrt{\bar{\beta}\bar{\alpha}PP_1}}{\alpha P + N_2}\right) + n\delta_{2,n}, \tag{101}$$

$$nR_0 + nR_2 < \frac{n}{2} \log\left(1 + \frac{\beta\bar{\alpha}P}{\alpha P + \frac{N_1 N_2}{N_1+N_2}}\right) + n\delta_{2,n}, \tag{102}$$

$$nR_1 < \frac{n}{2} \log\left(1 + \frac{\alpha P}{\frac{N_1 N_2}{N_1+N_2}}\right) + n\delta_{1,n}, \tag{103}$$

where parameters $\alpha$ and $\beta$ are defined by

$$\sum_{i=1}^{n} h(S_i|X_{1,i}, U_i) = \frac{n}{2} \log 2\pi e \left(\alpha P + \frac{N_1 N_2}{N_1 + N_2}\right)$$
$$\frac{1}{n} \sum_{i=1}^{n} \mathsf{E}\left(\mathsf{E}^2(X_i|X_{1,i})\right) = \bar{\beta}\bar{\alpha}P \tag{104}$$

We now further bound (100)

$$\begin{aligned}
nR_0 + nR_1 &\leq \sum_{i=1}^{n} I(U'_i; Y_{1,i}|X_{1,i}) + n\delta_{1,n} \\
&= \sum_{i=1}^{n} h(Y_{1,i}|X_{1,i}) - h(Y_{1,i}|U'_i, X_{1,i}) + n\delta_{1,n} \\
&\leq \sum_{i=1}^{n} \mathsf{E} h(X_i + Z_{1,i}|X_{1,i}) - h(Y_{1,i}|U'_i, X_{1,i}, X_i) + n\delta_{1,n} \\
&\leq \sum_{i=1}^{n} \mathsf{E} \frac{1}{2} \log 2\pi e \mathsf{Var}(X_i + Z_{1,i}|X_{1,i}) - \frac{n}{2} \log 2\pi e N_1 + n\delta_{1,n} \\
&\leq \frac{1}{2} \sum_{i=1}^{n} \log 2\pi e (\mathsf{E}\mathsf{Var}(X_i|X_{1,i}) + N_1) - \frac{n}{2} \log 2\pi e N_1 + n\delta_{1,n} \\
&\leq \frac{n}{2} \log 2\pi e \left( \frac{1}{n} \sum_{i=1}^{n} \mathsf{E}\mathsf{Var}(X_i|X_{1,i}) + N_1 \right) - \frac{n}{2} \log 2\pi e N_1 + n\delta_{1,n} \\
&= \frac{n}{2} \log 2\pi e \left( \frac{1}{n} \sum_{i=1}^{n} \mathsf{E}(X_i^2) - \mathsf{E}(\mathsf{E}^2(X_i|X_{1,i})) + N_1 \right) - \frac{n}{2} \log 2\pi e N_1 + n\delta_{1,n} \\
&= \frac{n}{2} \log 2\pi e \left( P - \bar{\beta}\bar{\alpha}P + N_1 \right) - \frac{n}{2} \log 2\pi e N_1 + n\delta_{1,n} \\
&= \frac{n}{2} \log \left( 1 + \frac{\alpha P + \beta \bar{\alpha} P}{N_1} \right) + n\delta_{1,n}
\end{aligned}$$
(105)

Therefore, the bounds given in (101)-(103) and (105) provide the outer bound given in Theorem 5.

## G   Outline of Proof for Theorem 10

We assume that the source uses the superposition coding [6, Chapter 14.6]. We also assume that user 1 employs the decode-and-forward relaying scheme [5, Section II], and user 2 employs the estimate-and-forward relaying scheme [5, Theorem 6]. As in Appendix A, we adopt the regular encoding/sliding window decoding strategy [2] for the decode-and-forward scheme. We also use the regular encoding/sliding window decoding strategy for the estimate-and-forward scheme, which is different from the irregular encoding/successive decoding strategy originally used in [5, Theorem 6].

We first prove that the rate region $\mathcal{R}_1$ is achievable for the case where $R_0 = 0$ (without common message $W_0$). It is then easily seen from the following proof that user 1 decodes the messages for both users 1 and 2. Hence we can always view part of the rate $R_2$ to be the common rate $R_0$. Therefore we will have proven that the rate region $\mathcal{R}_1$ given in Theorem 10 is achievable. The achievability of $\mathcal{R}_2$ can be shown by following steps similar to those used in proving the achievability of $\mathcal{R}_1$ with the roles of user 1 and user 2 being switched.

We consider a transmission over $B$ blocks, each with length $n$. At each of the first $B-2$ blocks, a message pair $(W_{1,i}, W_{2,i}) \in [1, 2^{nR_1}] \times [1, 2^{nR_2}]$ is encoded and sent from the source, where $i$ denotes the index of the block, and $i = 1, 2, \ldots, B-2$. For fixed $n$, the rate pair $\left(R_1 \frac{B-2}{B}, R_2 \frac{B-2}{B}\right)$ approaches $(R_1, R_2)$ as $B \to \infty$.

We use a random coding argument. Fix a probability distribution

$$p(x_1)p(u|x_1)p(x|u,x_1)p(x_2)p(y_1,y_2|x,x_1,x_2)p(\hat{y}_2|y_2,u,x_1,x_2), \tag{106}$$

and we use $A_\epsilon^{(n)}$ to denote the jointly $\epsilon$-typical set based on this joint distribution.

*Random Codebook Generation:* We generate three statistically independent random codebooks by following the same steps.

1. Generate $2^{nR_2}$ i.i.d. $\underline{x}_1$ each with distribution $\prod_{i=1}^n p(x_{1,i})$. Index $\underline{x}_1(w_2')$, $w_2' \in [1, 2^{nR_2}]$.

2. For each $\underline{x}_1(w_2')$, generate $2^{nR_2}$ i.i.d. $\underline{u}$ each with distribution $\prod_{i=1}^n p(u_i|x_{1,i}(w_2'))$. Index $\underline{u}(w_2', w_2)$, $w_2 \in [1, 2^{nR_2}]$.

3. For each $\underline{x}_1(w_2')$ and $\underline{u}(w_2', w_2)$, generate $2^{nR_1}$ i.i.d. $\underline{x}$ each with distribution $\prod_{i=1}^n p(x_i|u_i(w_2', w_2), x_{1,i}(w_2'))$. Index $\underline{x}(w_2', w_2, w_1)$, $w_1 \in [1, 2^{nR_1}]$.

4. Generate $2^{n\hat{R}_1}$ i.i.d. $\underline{x}_2$ each with distribution $\prod_{i=1}^n p(x_{2,i})$. Index $\underline{x}_2(z')$, $z' \in [1, 2^{n\hat{R}_1}]$.

5. For each $\underline{x}_1(w_2'), \underline{u}(w_2', w_2), \underline{x}_2(z')$, generate $2^{n\hat{R}_1}$ i.i.d. $\underline{\hat{y}}_2$ each with distribution $\prod_{i=1}^n p(\hat{y}_{2,i}|x_{1,i}(w_2'), u_i(w_2', w_2), x_{2,i}(z'))$, where the distribution $p(\hat{y}_2|u, x_1, x_2)$ is induced by the joint distribution given by (106). Index $\underline{\hat{y}}_2(w_2', w_2, z', z)$, $z \in [1, 2^{n\hat{R}_1}]$.

*Encoding:* We encode messages using codebooks 1, 2 and 3, respectively, for adjacent three blocks. This is because some of the following decoding steps are performed jointly over two or three adjacent blocks, and having independent codebooks makes the error events corresponding to these blocks independent, thus making the probabilities of these error events easy to calculate.

At the source, let $(w_{1,i}, w_{2,i})$ be the new message pair to be sent in current block $i$, and let $(w_{1,i-1}, w_{2,i-1})$ be the message pair being sent in previous block $i-1$. The source then sends $\underline{x}(w_{2,i-1}, w_{2,i}, w_{1,i})$.

At the beginning of block $i$, user 1 should have an estimation $\hat{w}_{2,i-1}$ of the message $w_{2,i-1}$ sent in the previous block $i-1$. It then sends $\underline{x}_1(\hat{w}_{2,i-1})$.

At the beginning of block $i$, user 2 should have an estimation $\hat{z}_{i-2}$ of the index $z_{i-2}$ of the compressed signal $\underline{\hat{y}}_2$. It then sends $\underline{x}_2(z_{i-2})$.

For convenience, we list the codewords sent in the first four blocks in the following table.

| block 1 | block 2 | block 3 | block 4 |
|---|---|---|---|
| $\underline{x}_1(1)$ | $\underline{x}_1(w_{2,1})$ | $\underline{x}_1(w_{2,2})$ | $\underline{x}_1(w_{2,3})$ |
| $\underline{u}\,(1, w_{2,1})$ | $\underline{u}\,(w_{2,1}, w_{2,2})$ | $\underline{u}\,(w_{2,2}, w_{2,3})$ | $\underline{u}\,(w_{2,3}, w_{2,4})$ |
| $\underline{x}\,(1, w_{2,1}, w_{1,1})$ | $\underline{x}\,(w_{2,1}, w_{2,2}, w_{1,2})$ | $\underline{x}\,(w_{2,2}, w_{2,3}, w_{1,3})$ | $\underline{x}\,(w_{2,3}, w_{2,4}, w_{1,4})$ |
| $\underline{x}_2(\qquad 1)$ | $\underline{x}_2(\qquad 1)$ | $\underline{x}_2(\qquad z_1)$ | $\underline{x}_2(\qquad z_2)$ |
| $\underline{\hat{y}}_2(1, w_{2,1}, 1, z_1)$ | $\underline{\hat{y}}_2(w_{2,1}, w_{2,2}, 1, z_2)$ | $\underline{\hat{y}}_2(w_{2,2}, w_{2,3}, z_1, z_3)$ | $\underline{\hat{y}}_2(w_{2,3}, w_{2,4}, z_2, z_4)$ |

*Decoding:* The decoding procedures at the end of block $i$ are as follows.

1. User 1, having known $w_{2,i-1}$, declares message $\hat{w}_{2,i}$ is sent if there is a unique $\hat{w}_{2,i}$ such that $\left(\underline{x}_1(w_{2,i-1}), \underline{u}(w_{2,i-1}, \hat{w}_{2,i}), \underline{y}_1(i)\right) \in A_\epsilon^{(n)}$. The decoding error in this step is small for sufficiently

large $n$ if
$$R_2 < I(U;Y_1|X_1). \tag{107}$$

2. User 1, having known $w_{2,i-3}, \ldots, w_{2,i}$ and $z_{i-4}$, determines that $\hat{\underline{y}}_2$ indexed by $\hat{z}_{i-2}$ is picked to compress $\underline{y}_2(i-1)$ by user 2 based on the information received in blocks $i-2$ and $i$. User 1 declares the index to be $\hat{z}_{i-2}$ if there is a unique $\hat{z}_{i-2}$ such that

$$\left(\underline{x}_1(w_{2,i-3}), \underline{u}(w_{2,i-3}, w_{2,i-2}), \underline{x}_2(z_{i-4}), \hat{\underline{y}}_2(w_{2,i-3}, w_{2,i-2}, z_{i-4}, \hat{z}_{i-2}), \underline{y}_1(i-2)\right) \in A_\epsilon^{(n)},$$

and $\quad \left(\underline{x}_1(w_{2,i-1}), \underline{u}(w_{2,i-1}, w_{2,i}), \underline{x}_2(\hat{z}_{i-2}), \underline{y}_1(i)\right) \in A_\epsilon^{(n)}.$

The decoding error in this step is small for sufficiently large $n$ if
$$\hat{R}_1 < I(\hat{Y}_2;Y_1|U,X_1,X_2) + I(X_2;Y_1|X_1,U). \tag{108}$$

3. User 1, having known $w_{2,i-3}, w_{2,i-2}$ and $z_{i-4}, z_{i-2}$, determines that the message $\hat{w}_{1,i-2}$ is sent based on the information received in block $i-2$. It declares the index to be $\hat{w}_{1,i-2}$ if there is a unique $\hat{w}_{1,i-2}$ such that

$$\left(\underline{x}_1(w_{2,i-3}), \underline{u}(w_{2,i-3}, w_{2,i-2}), \underline{x}(w_{2,i-3}, w_{2,i-2}, \hat{w}_{1,i-2}), \underline{x}_2(z_{i-4}),\right. \tag{109}$$
$$\left. \hat{\underline{y}}_2(w_{2,i-3}, w_{2,i-2}, z_{i-4}, z_{i-2}), \underline{y}_1(i-2)\right) \in A_\epsilon^{(n)}. \tag{110}$$

The decoding error in this step is small for sufficiently large $n$ if
$$R_1 < I(X;\hat{Y}_2,Y_1|U,X_1,X_2). \tag{111}$$

4. User 2, having known $w_{2,i-2}$, $z_{i-3}$ and $z_{i-2}$, determines that the message $\hat{\hat{w}}_{2,i-1}$ is sent based on the information received in blocks $i-1$ and $i$. It declares the index to be $\hat{\hat{w}}_{2,i-1}$ if there is a unique $\hat{\hat{w}}_{2,i-1}$ such that

$$\left(\underline{x}_1(w_{2,i-2}), \underline{u}(w_{2,i-2}, \hat{\hat{w}}_{2,i-1}), \underline{x}_2(z_{i-3}), \underline{y}_2(i-1)\right) \in A_\epsilon^{(n)},$$

and $\quad \left(\underline{x}_1(\hat{\hat{w}}_{2,i-1}), \underline{x}_2(z_{i-2}), \underline{y}_2(i)\right) \in A_\epsilon^{(n)}.$

The decoding error in this step is small for sufficiently large $n$ if
$$R_2 < I(U;Y_2|X_1,X_2) + I(X_1;Y_2|X_2) = I(X_1,U;Y_2|X_2). \tag{112}$$

5. User 2, having known $w_{2,i-2}, w_{2,i-1}$ and $z_{i-3}$, declares that the estimate signal $\hat{\underline{y}}_2$ for $\underline{y}_2(i-1)$ is indexed by $\hat{\hat{z}}_{i-1}$ if there is a unique $\hat{\hat{z}}_{i-1}$ such that

$$\left(\underline{x}_1(w_{2,i-2}), \underline{u}(w_{2,i-2}, w_{2,i-1}), \underline{x}_2(z_{i-3}), \hat{\underline{y}}_2(w_{2,i-2}, w_{2,i-1}, z_{i-3}, \hat{\hat{z}}_{i-1}), \underline{y}_2(i-1)\right) \in A_\epsilon^{(n)}.$$

There exists such a $z_{i-1}$ with high probability for sufficiently large $n$ if
$$\hat{R}_1 > I(\hat{Y}_2;Y_2|U,X_1,X_2). \tag{113}$$

Combining (108) and (113), we obtain
$$\begin{aligned}I(X_2;Y_1|X_1,U) &> I(\hat{Y}_2;Y_2|U,X_1,X_2) - I(\hat{Y}_2;Y_1|U,X_1,X_2) \\ &= I(\hat{Y}_2;Y_2|Y_1,U,X_1,X_2),\end{aligned} \tag{114}$$

which is exactly the constraint given in $\mathcal{R}_1$.

Combining (107), (111), (112), and (114), we obtain the rate region $\mathcal{R}_1$.

# H   Outline of Proof for Theorem 14

Let $\hat{Y}_2 = Y_2 + \check{Z}$ where the variance of $\check{Z}$ is denoted by $\check{N}$ that will be determined later in the proof.

We compute the achievable rate region $\mathcal{R}_1$ given in Theorem 10 based on the following distributions and relationships for those random variables in the expression of $\mathcal{R}_1$

$$
\begin{aligned}
&X_1 \sim \mathcal{N}(0, P_1), \\
&U' \sim \mathcal{N}(0, \beta\bar{\alpha}P), \qquad U = \sqrt{\tfrac{\bar{\beta}\bar{\alpha}P}{P_1}} X_1 + U', \\
&X' \sim \mathcal{N}(0, \alpha P), \qquad X = U + X', \\
&X_2 \sim \mathcal{N}(0, \eta P_2),
\end{aligned}
\tag{115}
$$

where the random variables $X_1, U', X', X_2$ are independent.

It is straightforward to compute the following two mutual information terms that provide the expression for $R_0 + R_2$

$$
\begin{aligned}
I(U, X_1; Y_2 | X_2) &= \mathcal{C}\left(\frac{\bar{\alpha}P + P_1 + 2\sqrt{\bar{\beta}\bar{\alpha}PP_1}}{\alpha P + N_2}\right), \\
I(U; Y_1 | X_1) &= \mathcal{C}\left(\frac{\beta\bar{\alpha}P}{\alpha P + +\eta P_2 + N_1}\right).
\end{aligned}
\tag{116}
$$

To derive $R_1$, we first have

$$
R_1 < I(X; \hat{Y}_2, Y_1 | X_1, U, X_2) = \mathcal{C}\left(\frac{\alpha P}{N_1} + \frac{\alpha P}{N_2 + \check{N}}\right). \tag{117}
$$

To determine $\check{N}$ in the preceding equation, we use the following constraint which is given in the expression of $\mathcal{R}_1$

$$
\begin{aligned}
I(X_2; Y_1 | U, X_1) &\geq I(\hat{Y}_2; Y_2 | Y_1, U, X_1, X_2) \\
&= I(\hat{Y}_2; Y_2 | U, X_1, X_2) - I(\hat{Y}_2; Y_1 | U, X_1, X_2).
\end{aligned}
\tag{118}
$$

We evaluate the mutual information terms in (118), and have

$$
\begin{aligned}
I(X_2; Y_1 | U, X_1) &= \tfrac{1}{2} \log\left(1 + \frac{\eta P_2}{\alpha P + N_1}\right), \\
I(\hat{Y}_2; Y_2 | U, X_1, X_2) &= \tfrac{1}{2} \log\left(1 + \frac{\alpha P + N_2}{\check{N}}\right), \\
I(\hat{Y}_2; Y_1 | U, X_1, X_2) &= \tfrac{1}{2} \log\left(\frac{\alpha P + N_2 + \check{N}}{\alpha P + N_2 + \check{N} - \frac{(\alpha P)^2}{\alpha P + N_1}}\right).
\end{aligned}
\tag{119}
$$

We plug the three mutual information terms given in (119) into (118), and derive the following constraint on $\check{N}$

$$
\check{N} \geq \frac{\alpha P(N_1 + N_2) + N_1 N_2}{\eta P_2}. \tag{120}
$$

We now plug the preceding bound on $\check{N}$ in the expression (117), and obtain

$$
R_1 < \mathcal{C}\left(\frac{\alpha P}{N_1} + \frac{\alpha \eta PP_2}{\eta P_2 N_2 + \alpha P(N_1 + N_2) + N_1 N_2}\right). \tag{121}
$$

# I Outline of Proof for Theorem 15

In the proof for Theorem 18, we have shown that the mapping from $(Y_1, Y_2)$ to $(S, Y_2)$ is one-to-one, where

$$S = X + \frac{N_1}{N_1 + N_2}X_1 + \frac{N_2}{N_1 + N_2}X_2 + \hat{Z}_1 \tag{122}$$
$$Y_2 = X + X_1 + \hat{Z}_1 + \hat{Z}$$

where $\hat{Z}_1$ and $\hat{Z}$ are independent zero mean real Gaussian random variables with variances $\frac{N_1 N_2}{N_1 + N_2}$ and $\frac{N_2^2}{N_2 + N_1}$, respectively.

We define the following two auxiliary random variables:

$$\begin{aligned} U_i &:= (W_0, W_2, Y_1^{i-1}, Y_2^{i-2}), \\ U_i' &:= (W_0, W_1, Y_1^{i-1}, Y_2^{i-2}). \end{aligned} \tag{123}$$

We follow the steps similar to those in Appendix B, and derive the following upper bounds

$$nR_0 + nR_2 = \sum_{i=1}^n I(U_i, X_{1,i}; Y_{2,i}|X_{2,i}) + n\delta_{2,n} \tag{124}$$

$$\begin{aligned} nR_0 + nR_2 &\leq \sum_{i=1}^n I(U_i; Y_{1,i}, Y_{2,i}|X_{1,i}, X_{2,i}) + n\delta_{2,n} = \sum_{i=1}^n I(U_i; S_i, Y_{2,i}|X_{1,i}, X_{2,i}) + n\delta_{2,n} \\ &= \sum_{i=1}^n I(U_i; S_i|X_{1,i}, X_{2,i}) + n\delta_{2,n} \end{aligned} \tag{125}$$

$$\begin{aligned} nR_1 &\leq \sum_{i=1}^n I(X_i; Y_{1,i}, Y_{2,i}|U_i, X_{1,i}, X_{2,i}) + n\delta_{1,n} = \sum_{i=1}^n I(X_i; S_i, Y_{2,i}|U_i, X_{1,i}, X_{2,i}) + n\delta_{1,n} \\ &= \sum_{i=1}^n I(X_i; S_i|U_i, X_{1,i}, X_{2,i}) + n\delta_{1,n} \end{aligned} \tag{126}$$

$$nR_0 + nR_1 \leq \sum_{i=1}^n I(U_i', X_{2,i}; Y_{1,i}|X_{1,i}) + n\delta_{1,n} \tag{127}$$

$$\begin{aligned} nR_0 + nR_1 &\leq \sum_{i=1}^n I(U_i'; Y_{1,i}, Y_{2,i}|X_{1,i}, X_{2,i}) + n\delta_{1,n} = \sum_{i=1}^n I(U_i'; S_i, Y_{2,i}|X_{1,i}, X_{2,i}) + n\delta_{1,n} \\ &= \sum_{i=1}^n I(U_i'; S_i|X_{1,i}, X_{2,i}) + n\delta_{1,n} \end{aligned} \tag{128}$$

$$\begin{aligned} nR_2 &\leq \sum_{i=1}^n I(X_i; Y_{1,i}, Y_{2,i}|U_i', X_{1,i}, X_{2,i}) + n\delta_{2,n} = \sum_{i=1}^n I(X_i; S_i, Y_{2,i}|U_i', X_{1,i}, X_{2,i}) + n\delta_{2,n} \\ &= \sum_{i=1}^n I(X_i; S_i|U_i', X_{1,i}, X_{2,i}) + n\delta_{2,n} \end{aligned} \tag{129}$$

where for equations (125), (126), (128), and (129), we have used the fact that the mapping from $(Y_1, Y_2)$ to $(S, Y_2)$ is one-to-one and the fact that given $(S_i, X_{1,i}, X_{2,i})$, $Y_{2,i}$ is independent of $U_i$ and $X_i$.

We further bound equations (124), (125) and (126) by following similar steps in bounding equations (75), (76) and (77) in Appendix E with $Y_1$ being replaced by $S$ and $N_1$ being replaced by $\frac{N_1 N_2}{N_1+N_2}$. We then obtain the following bounds

$$nR_0 + nR_2 < \frac{n}{2}\log\left(1 + \frac{P_1 + \bar{\alpha}P + 2\sqrt{\bar{\beta}\bar{\alpha}PP_1}}{\alpha P + N_2}\right) + n\delta_{2,n},$$

$$nR_0 + nR_2 < \frac{n}{2}\log\left(1 + \frac{\beta\bar{\alpha}P}{\alpha P + \frac{N_1 N_2}{N_1+N_2}}\right) + n\delta_{2,n}, \quad (130)$$

$$nR_1 < \frac{n}{2}\log\left(1 + \frac{\alpha P}{\frac{N_1 N_2}{N_1+N_2}}\right) + n\delta_{1,n}$$

where parameters $\alpha$ and $\beta$ are defined by

$$\sum_{i=1}^{n} h(S_i|X_{1,i}, X_{2,i}, U_i) = \frac{n}{2}\log 2\pi e\left(\alpha P + \frac{N_1 N_2}{N_1 + N_2}\right)$$

$$\frac{1}{n}\sum_{i=1}^{n} \mathsf{E}\left(\mathsf{E}^2(X_i|X_{1,i}, X_{2,i})\right) = \bar{\beta}\bar{\alpha}P \quad (131)$$

We also obtain the following intermediate bound which will be useful for the rest of the proof

$$\sum_{i=1}^{n} h(S_i|X_{1,i}, X_{2,i}) \leq \frac{n}{2}\log 2\pi e\left(\alpha P + \beta\bar{\alpha}P + \frac{N_1 N_2}{N_1 + N_2}\right) \quad (132)$$

We now further bound (127)

$$nR_0 + nR_1 \leq \sum_{i=1}^{n} I(U'_i, X_{2,i}; Y_{1,i}|X_{1,i}) + n\delta_{1,n}$$

$$= \sum_{i=1}^{n} h(Y_{1,i}|X_{1,i}) - h(Y_{1,i}|U'_i, X_{1,i}, X_{2,i}) + n\delta_{1,n}$$

$$\leq \sum_{i=1}^{n} h(X_i + X_{2,i} + Z_{1,i}) - h(Y_{1,i}|U'_i, X_{1,i}, X_{2,i}, X_i) + n\delta_{1,n} \quad (133)$$

$$\leq \sum_{i=1}^{n} \frac{1}{2}\log 2\pi e(\mathsf{E}(X_i + X_{2,i})^2 + N_1) - \frac{n}{2}\log 2\pi e N_1 + n\delta_{1,n}$$

$$\leq \frac{n}{2}\log 2\pi e\left(\frac{1}{n}\sum_{i=1}^{n}\mathsf{E}(X_i + X_{2,i})^2 + N_1\right) - \frac{n}{2}\log 2\pi e N_1 + n\delta_{1,n}$$

For the sum in the preceding equation, we have

$$\frac{1}{n}\sum_{i=1}^{n}\mathsf{E}(X_i+X_{2,i})^2 = \frac{1}{n}\sum_{i=1}^{n}\mathsf{E}X_i^2 + \frac{1}{n}\sum_{i=1}^{n}\mathsf{E}X_{2,i}^2 + \frac{2}{n}\sum_{i=1}^{n}\mathsf{E}X_iX_{2,i}$$

$$\leq P + P_2 + \frac{2}{n}\sum_{i=1}^{n}\mathsf{E}\left(X_{2,i}\mathsf{E}(X_i|X_{1,i},X_{2,i})\right)$$

$$\leq P + P_2 + \frac{2}{n}\sum_{i=1}^{n}\sqrt{\mathsf{E}X_{2,i}^2 \cdot \mathsf{E}\left(\mathsf{E}^2(X_i|X_{1,i},X_{2,i})\right)} \tag{134}$$

$$\leq P + P_2 + 2\sqrt{\left(\frac{1}{n}\sum_{i=1}^{n}\mathsf{E}X_{2,i}^2\right)\cdot\left(\frac{1}{n}\sum_{i=1}^{n}\mathsf{E}\left(\mathsf{E}^2(X_i|X_{1,i},X_{2,i})\right)\right)}$$

$$\leq P + P_2 + 2\sqrt{P_2\bar{\beta}\bar{\alpha}P}$$

Hence we have,

$$nR_0 + nR_1 \leq \frac{n}{2}\log\left(1 + \frac{P + P_2 + 2\sqrt{\bar{\beta}\bar{\alpha}PP_2}}{N_1}\right) + n\delta_{1,n} \tag{135}$$

We next bound (128)

$$nR_0 + nR_1$$
$$\leq \sum_{i=1}^{n} I(U_i'; S_i|X_{1,i},X_{2,i}) + n\delta_{1,n}$$
$$= \sum_{i=1}^{n} h(S_i|X_{1,i},X_{2,i}) - h(S_i|U_i',X_{1,i},X_{2,i}) + n\delta_{1,n}$$
$$\leq \frac{n}{2}\log 2\pi e\left(\alpha P + \beta\bar{\alpha}P + \frac{N_1N_2}{N_1+N_2}\right) - \frac{n}{2}\log 2\pi e\left(\gamma(\alpha P + \beta\bar{\alpha}P) + \frac{N_1N_2}{N_1+N_2}\right) + n\delta_{1,n}$$
$$\leq \frac{n}{2}\log 2\pi e\left(1 + \frac{\bar{\gamma}(\alpha P + \beta\bar{\alpha}P)}{\gamma(\alpha P + \beta\bar{\alpha}P) + \frac{N_1N_2}{N_1+N_2}}\right) + n\delta_{1,n}$$
$$\tag{136}$$

where we used the fact that there exists a $\gamma \in [0,1]$ such that

$$\sum_{i=1}^{n} h(S_i|U_i',X_{1,i},X_{2,i}) = \frac{n}{2}\log 2\pi e\left(\gamma(\alpha P + \beta\bar{\alpha}P) + \frac{N_1N_2}{N_1+N_2}\right). \tag{137}$$

The preceding equation follows from

$$\sum_{i=1}^{n} h(S_i|U_i',X_{1,i},X_{2,i}) \leq \sum_{i=1}^{n} h(S_i|X_{1,i},X_{2,i}) \leq \frac{n}{2}\log 2\pi e\left(\alpha P + \beta\bar{\alpha}P + \frac{N_1N_2}{N_1+N_2}\right), \tag{138}$$

and

$$\sum_{i=1}^{n} h(S_i|U_i',X_{1,i},X_{2,i}) \geq \sum_{i=1}^{n} h(S_i|U_i',X_{1,i},X_{2,i},X_i) = \frac{n}{2}\log 2\pi e\frac{N_1N_2}{N_1+N_2}. \tag{139}$$

We finally bound (129)

$$\begin{aligned}
nR_2 &\leq \sum_{i=1}^{n} I(X_i; S_i | U'_i, X_{1,i}, X_{2,i}) + n\delta_{2,n} \\
&= \sum_{i=1}^{n} h(S_i | U'_i, X_{1,i}, X_{2,i}) - h(S_i | U'_i, X_{1,i}, X_{2,i}, X_i) + n\delta_{2,n} \\
&= \frac{n}{2} \log 2\pi e \left( \gamma(\alpha P + \beta\bar{\alpha}P) + \frac{N_1 N_2}{N_1 + N_2} \right) - \frac{n}{2} \log 2\pi e \frac{N_1 N_2}{N_1 + N_2} + n\delta_{2,n} \\
&= \frac{n}{2} \log 2\pi e \left( 1 + \frac{\gamma(\alpha P + \beta\bar{\alpha}P)}{\frac{N_1 N_2}{N_1 + N_2}} \right) + n\delta_{2,n}
\end{aligned} \qquad (140)$$

# Acknowledgement

The authors would like to thank Dr. Gerhard Kramer at Bell Laboratories of Lucent Technologies for useful discussions.

# References

[1] K. Azarian, H. El Gamal, and P. Schniter. On the achievable diversity-multiplexing tradeoffs in half-duplex cooperative channels. submitted to *IEEE Trans. Inform. Theory*, July 2004.

[2] A. B. Carleial. Multiple-access channels with different generalized feedback signals. *IEEE Trans. Inform. Theory*, 28(6):841–850, November 1982.

[3] T. M. Cover. Broadcast channels. *IEEE Trans. Inform. Theory*, 18(1):2–14, January 1972.

[4] T. M. Cover. Comments on broadcast channels. *IEEE Trans. Inform. Theory*, 44(6):2524–2530, October 1998.

[5] T. M. Cover and A. A. El Gamal. Capacity theorems for the relay channel. *IEEE Trans. Inform. Theory*, 25(5):572–584, September 1979.

[6] T. M. Cover and J. A. Thomas. *Elements of Information Theory*. Wiley, New York, 1991.

[7] I. Csiszár and J. Körner. *Information Theory: Coding Theorems for Discrete Memoryless Systems*. Akadémiai Kiadó, Budapest, 1981.

[8] R. Dabora and S. Servetto. Broadcast channels with cooperating decoders. submitted to *IEEE Trans. Inform. Theory*, May 2005.

[9] R. Dabora and S. Servetto. Broadcast channels with cooperating receivers: a downlink for the sensor reachback problem. In *Proc. IEEE Int. Symp. Inform. Theory (ISIT)*, page 176, Chicago, IL, June-July 2004.

[10] A. A. El Gamal. The feedback capacity of degraded broadcast channels. *IEEE Trans. Inform. Theory*, 24(3):379–381, May 1978.

[11] A. A. El Gamal. The capacity of the physically degraded Gaussian broadcast channel with feedback. *IEEE Trans. Inform. Theory*, 27(4):508–511, July 1981.

[12] A. A. El Gamal and S. Zahedi. Capacity of a class of relay channels with orthogonal components. *IEEE Trans. Inform. Theory*, 51(5):1815–1817, May 2005.

[13] A. El Gamal, M. Mohseni, and S. Zahedi. On reliable communication over additive white Gaussian noise relay channels. submitted to *IEEE Trans. Inform. Theory*, 2004.

[14] M. Gastpar and M. Vetterli. On the capacity of large Gaussian relay networks. *IEEE Trans. Inform. Theory*, 51(3):765–779, March 2005.

[15] D. Gunduz and E. Erkip. Outage minimization by opportunistic cooperation. In *Proc. WirelessCom, Symp. on Inform. Th.*, Maui, Hawaii, June 2005.

[16] P. Gupta and P. R. Kumar. Towards an information theory of large networks: an achievable rate region. *IEEE Trans. Inform. Theory*, 49(8):1877–1894, August 2003.

[17] A. Host-Madsen. On the achievable rate for receiver cooperation in ad-hoc networks. In *Proc. IEEE Int. Symp. Inform. Theory (ISIT)*, page 272, Chicago, June-July 2004.

[18] A. Host-Madsen and J. Zhang. Capacity bounds and power allocation for wireless relay channels. *IEEE Trans. Inform. Theory*, 51(6):2020–2040, June 2005.

[19] N. Jindal, U. Mitra, and A. Goldsmith. Capacity of ad-hoc networks with node cooperation. In *Proc. IEEE Int. Symp. Inform. Theory (ISIT)*, page 271, Chicago, June-July 2004.

[20] M. Katz and S. Shamai (Shitz). Communicating to co-located ad-hoc receiving nodes in a fading environment. In *Proc. IEEE Int. Symp. Inform. Theory (ISIT)*, page 115, Chicago, June-July 2004.

[21] M. A. Khojastepour and B. Aazhang. 'Cheap' relay channel: a unifying approach to time and frequency division relaying. In *Proc. Annual Allerton Conf. on Communication, Control and Computing*, pages 1792–1801, Monticello, Illinois, Sept.-Oct. 2004.

[22] M. A. Khojastepour, A. Sabharwal, and B. Aazhang. Lower bounds on the capacity of Gaussian relay channel. In *Proc. Conf. on Information Sciences and Systems (CISS)*, pages 597–602, Princeton, New Jersey, March 2004.

[23] G. Kramer. Models and theory for relay channels with receiver constraints. In *Proc. Annual Allerton Conf. on Communication, Control and Computing*, pages 1312–1321, Monticello, Illinois, Sept.-Oct. 2004.

[24] G. Kramer, M. Gastpar, and P. Gupta. Cooperative strategies and capacity theorems for relay networks. To appear in *IEEE Trans. Inform. Theory*, 2005.

[25] G. Kramer and A. J. van Wijngaarden. On the white Gaussian multiple-access relay channel. In *Proc. IEEE Int. Symp. Inform. Theory (ISIT)*, page 40, Sorrento, Italy, June 2000.

[26] J. N. Laneman, D. N. C. Tse, and G. W. Wornell. Cooperative diversity in wireless networks: efficient protocols and outage behavior. *IEEE Trans. Inform. Theory*, 50(12):3062–3080, December 2004.

[27] J. N. Laneman and G. W. Wornell. Distributed space-time coded protocols for exploring cooperative diversity in wireless networks. *IEEE Trans. Inform. Theory*, 49(10):2415–2425, October 2003.

[28] Y. Liang and V. V. Veeravalli. Guassian orthogonal relay channels: optimal resource allocation and capacity. To appear in *IEEE Trans. Inform. Th.*, 2005.

[29] Y. Liang and V. V. Veeravalli. The impact of relaying on the capacity of broadcast channels. In *Proc. IEEE Int. Symp. Inform. Theory (ISIT)*, page 403, Chicago, June-July 2004.

[30] Y. Liang and V. V. Veeravalli. Resource allocation for wireless relay channels. In *Proc. Asilomar Conf. on Signals, Systems and Computers*, Pacific Grove, California, November 2004.

[31] Y. Liang and V. V. Veeravalli. Distributed optimal resource allocation for fading relay broadcast channels. In *IEEE International Workshop on Signal Processing Advances for Wireless Communications (SPAWC)*, New York, June 2005.

[32] I. Maric and R. Yates. Forwarding strategies for gaussian parallel-relay networks. In *Proc. IEEE Int. Symp. Inform. Theory (ISIT)*, page 269, Chicago, June-July 2004.

[33] I. Maric and R. D. Yates. Bandwidth and power allocation for cooperative strategies in Gaussian relay networks. In *Proc. Asilomar Conf. on Signals, Systems and Computers*, Pacific Grove, California, November 2004.

[34] U. Mitra and A. Sabharwal. On achievable rates of complexity constrained relay channels. In *Proc. Annual Allerton Conf. on Communication, Control and Computing*, pages 551–560, Monticello, Illinois, Sept.-Oct. 2003.

[35] R. U. Nabar, H. Bölcskei, and F. W. Kneubuhler. Fading relay channels: performance limits and space-time signal design. *IEEE J. Sel. Areas Commun.*, 22(6):1099–1109, August 2004.

[36] L. Ozarow. *Coding and capacity for additive white Gaussian noise multi-user channels with feedback.* Ph.D. dissertation, Department of Electrical Engineering and Computer Science, MIT, Cambridge, MA, 1983.

[37] L. H. Ozarow and S. K. Leung-Yan-Cheong. An achievable region and outer bound for the Gaussian broadcast channel with feedback. *IEEE Trans. Inform. Theory*, 30(4):667–671, 1984.

[38] N. Pradad and M. K. Varanasi. Diversity and multiplexing tradeoff bounds for cooperative diversity protocols. In *Proc. IEEE Int. Symp. Inform. Theory (ISIT)*, page 268, Chicago, June-July 2004.

[39] A. Reznik, S. Kulkarni, and S. Verdu. Degraded Gaussian multirelay channel: capacity and optimal power allocation. *IEEE Trans. Inform. Theory*, 50(12):3037–3046, December 2004.

[40] L. Sankaranarayanan, G. Kramer, and N. B. Mandayam. Capacity theorems for the multiple-access relay channel. In *Proc. Annual Allerton Conf. on Communication, Control and Computing*, pages 1782–1791, Monticello, Illinois, Sept.-Oct. 2004.

[41] L. Sankaranarayanan, G. Kramer, and N. B. Mandayam. Hierarchical wireless networks: capacity bounds using the constrained multiple-access relay channel model. In *Proc. Asilomar Conf. on Signals, Systems and Computers*, Pacific Grove, California, November 2004.

[42] B. Schein and R. Gallager. The Gaussian parallel relay network. In *Proc. IEEE Int. Symp. Inform. Theory (ISIT)*, page 22, Sorrento, Italy, June 2000.

[43] A. Sendonaris, E. Erkip, and B. Aazhang. User cooperation diversity-part I: system description. *IEEE Trans. Commun.*, 51(11):1927–1938, November 2003.

[44] A. Sendonaris, E. Erkip, and B. Aazhang. User cooperation diversity-part II: implementation aspects and performance analysis. *IEEE Trans. Commun.*, 51(11):1939–1948, November 2003.

[45] A. Stefanov and E. Erkip. Cooperative coding for wireless networks. *IEEE Trans. Commun.*, 52(9):1470–1476, September 2004.

[46] E. C. van der Meulen. Three-terminal communication channels. *Adv. Appl. Prob.*, 3:120–154, 1971.

[47] B. Wang, J. Zhang, and A. Host-Madsen. On the capacity of MIMO relay channels. *IEEE Trans. Inform. Theory*, 51(1):29–43, January 2005.

[48] L.-L. Xie and P. R. Kumar. A network information theory for wireless communication: scaling laws and optimal operation. *IEEE Trans. Inform. Theory*, 50(5):748–767, May 2004.

[49] L.-L. Xie and P. R. Kumar. An achievable rate for the multiple-level relay channel. *IEEE Trans. Inform. Theory*, 51(4):1348–1358, April 2005.

[50] Y. Yao, X. Cai, and G. B. Giannakis. On energy efficiency and optimum resource allocation in wireless relay transmissions. To appear in *IEEE Transactions on Wireless Communications*, 2005.